\documentclass[letterpaper,twocolumn,10pt]{article}
\usepackage{usenix2019_v3}

\usepackage[english]{babel}
\usepackage[font=footnotesize]{caption}
\usepackage[font=footnotesize]{subcaption}
\usepackage{fancybox}
\usepackage[frozencache]{minted}
\usepackage[babel]{csquotes}
\usepackage{comment}

\usepackage{listings}

\usepackage{colortbl} %
\usepackage{amsmath}    
\usepackage{amssymb}                                                        
\usepackage{amsthm, thmtools, amsfonts} %
\usepackage{tikz}
\usetikzlibrary{positioning,arrows.meta}
\usepackage[linesnumbered,ruled,vlined]{algorithm2e}
\usepackage{enumitem}
\usepackage{multirow}
\usepackage[olditem,oldenum]{paralist}
\usepackage{tabularx}
\usepackage{booktabs}
\usepackage[flushleft]{threeparttable}

\usepackage{pmboxdraw} %

\usepackage{pifont} %
\usepackage{tablefootnote}
\usepackage{xspace}
\usepackage{pgfplots} 
\usepackage{graphicx}
\usepackage{zi4} %

\pgfplotsset{compat=1.15}

\definecolor{sangria}{rgb}{0.57, 0.0, 0.04}
\definecolor{pinegreen}{rgb}{0.0, 0.47, 0.44}
\definecolor{rossocorsa}{rgb}{0.83, 0.0, 0.0}
\definecolor{ao}{rgb}{0.0, 0.0, 1.0}

\newcommand{\etal}{et~al.\xspace}
\newcommand{\eg}{e.\,g.,\xspace}
\newcommand{\ie}{i.\,e.,\xspace}
\newcommand{\cf}{cf.\xspace}

\newcommand{\inlinecode}[1]{{\tt #1}}

\newcommand{\ex}[1]{Experiment~\ref{#1}}

\newcolumntype{R}{>{\raggedleft\arraybackslash}X}

\newtheoremstyle{customexperiment}
  {0.25em} %
  {0.25em} %
  {\itshape}  %
  {0pt}       %
  {\bfseries} %
  {.}         %
  {5pt plus 1pt minus 1pt} %
  {\thmname{#1}\thmnumber{ #2}:\thmnote{ #3}}          %

\newtheoremstyle{customexample}
  {0.2em} %
  {0.2em} %
  {\itshape}  %
  {0pt}       %
  {\bfseries} %
  {}         %
  {4pt plus 1pt minus 1pt} %
  {\thmname{#1}\thmnumber{ #2}:\thmnote{ #3}}          %

\theoremstyle{customexample}
\newtheorem{example}{Example}

\theoremstyle{customexperiment}
\newtheorem{experiment}{Experiment}

\newcommand{\loki}{\textsc{Loki}\xspace}
\newcommand{\lokiattack}{\textsc{LokiAttack}\xspace}

\newcommand{\vmhunt}{\textsc{VMHunt}\xspace}
\newcommand{\vmattack}{\textsc{VMAttack}\xspace}
\newcommand{\neureduce}{\textsc{NeuReduce}\xspace}
\newcommand{\sspam}{\textsc{SSPAM}\xspace}
\newcommand{\arybo}{\textsc{arybo}\xspace}
\newcommand{\qsynth}{\textsc{QSynth}\xspace}
\newcommand{\mbablast}{\textsc{MBA-Blast}\xspace}
\newcommand{\xyntia}{\textsc{Xyntia}\xspace}
\newcommand{\syntia}{\textsc{Syntia}\xspace}
\newcommand{\tigress}{\textsc{Tigress}\xspace}

\newcommand{\themida}{\textsc{Themida}\xspace}
\newcommand{\vmprotect}{\textsc{VMProtect}\xspace}

\newcommand{\miasm}{\textsc{Miasm}\xspace}
\newcommand{\triton}{\textsc{Triton}\xspace}
\newcommand{\zt}{\textsc{Z3}\xspace}
\newcommand{\llvm}{{\textsc{LLVM}}\xspace}

\newcommand{\vlc}{\textsc{VLC}\xspace}
\newcommand{\libdvdcss}{\textsc{libdvdcss}\xspace}

\begin{document}

\date{}

\title{\Large \bf Technical Report: Hardening Code Obfuscation Against Automated Attacks}

\author{
{\rm Moritz Schloegel$^\star$, Tim Blazytko$^\star$, Moritz Contag$^\star$, Cornelius Aschermann$^\star$} \\%
{\rm Julius Basler$^\star$, Thorsten Holz$^\dagger$, Ali Abbasi$^\star$}\\\\%
$^\star$Ruhr-Universit\"{a}t Bochum\\
$^\dagger$CISPA Helmholtz Center for Information Security\\
} %

\maketitle{}

\vspace{-5em}
\begin{abstract}
\emph{Software obfuscation} is a crucial technology to protect intellectual property and manage digital rights within our society. Despite its huge practical importance, both commercial and academic state-of-the-art obfuscation methods are vulnerable to a plethora of automated deobfuscation attacks, such as symbolic execution, taint analysis, or program synthesis. While several enhanced obfuscation techniques were recently proposed to thwart taint analysis or symbolic execution, they either impose a prohibitive runtime overhead or can be removed in an automated way (\eg via compiler optimizations). In general, these techniques suffer from focusing on a single attack vector, allowing an attacker to switch to other, more effective techniques, such as program synthesis.

In this work, we present \loki, an approach for software obfuscation that is resilient against all known automated deobfuscation attacks. To this end, we use and efficiently combine multiple techniques, including a generic approach to synthesize formally verified expressions of arbitrary complexity. Contrary to state-of-the-art approaches that rely on a few hardcoded generation rules, our expressions are more diverse and harder to pattern match against. Even the most recent state-of-the-art research on Mixed-Boolean Arithmetic (MBA) deobfuscation fails to simplify them. Moreover, \loki protects against previously unaccounted attack vectors such as program synthesis, for which it reduces the success rate to merely 19\%. In a comprehensive evaluation, we show that our design incurs significantly less overhead while providing a much stronger protection level compared to existing works.
\end{abstract}

\section{Introduction}\label{sec:intro}

\emph{Obfuscation} describes the process of applying transformations to a given program with the goal of protecting the code from prying eyes. Generally speaking, obfuscation works by taking (parts of) a program and transforming it into a more complex, less intelligible representation, while at the same time preserving its observable input-output behavior~\cite{collberg1997taxonomy}. Usually, such transformations come at the cost of increased program runtime and size, thus trading intelligibility for overhead. Although formal verification of code transformations is hard to achieve in practice~\cite{lu2020formal,yang2011finding}, obfuscation is used in a wide range of real-world scenarios. Examples include protection of intellectual property (IP), digital rights management (DRM), and concealment of malicious behavior in software. Generally speaking, obfuscation protects critical (often small) code parts against reverse engineering and, thus, misuse by competitors or other parties. For example, most contemporary DRM systems rely on some kind of obfuscation to prevent attackers from distributing unauthorized copies of their product~\cite{nakanishi2020intertwining}. License checks and cryptographic authentication schemes are examples for code that is commonly obfuscated in practice to prevent analysis. Most copy-protection schemes used by games use some kind of obfuscation to prevent unauthorized copies. As another example, market-leading companies, such as \emph{Snapchat}, obfuscate how API calls to their backend are constructed, preventing abuse and access by competitors~\cite{snapchat_obfuscation}. %

Among the countless obfuscation methods proposed in the literature~\cite{collberg1997taxonomy,zhou2007information,obf-general,obf-probfuscation,obf-vmbased,obf-controlflow,obf-selfmodify,banescu2016code,wang2011linear,ollvm,xu2018manufacturing,ollivier2019howto,borrello2021hiding,suk2020vcf,xue2018exploiting,cheng2019dynopvm}, one of the most promising techniques is Virtual Machine (VM)-based obfuscation~\cite{rolles2009unpacking,obf-vmbased}. State-of-the-art, commercial obfuscators such as \themida~\cite{themida} and \vmprotect~\cite{vmprotect}, as well as most game copy-protection schemes used in practice~\cite{securom,denuvo}, make extensive use of VM-based obfuscation. They transform the to-be-protected code from its original Instruction Set Architecture (ISA) into a \emph{custom} one, and bundle an interpreter with the program that emulates the new ISA. This effectively breaks any analysis tool unfamiliar with the new architecture. Attackers aiming to deobfuscate code affected by this scheme must first uncover the custom ISA before they can reconstruct the original code~\cite{rolles2009unpacking,sharif2009automatic}.
Since the custom instruction sets are conceptually simple, VM-based obfuscation software usually applies additional obfuscating transformations to the interpreter such that it is harder to analyze. 
For example, dead code insertion or constant unfolding are often used.
At their core, these transformations inflate the number of executed instructions and mainly add to the code's \emph{syntactic} complexity, but can be successful in thwarting manual attacks.

However, it is often sufficient to apply well-known compiler optimizations, such as \emph{dead code elimination}, \emph{constant folding}, or \emph{constant propagation}, to reduce the code's \emph{syntactic} complexity and enable subsequent analyses~\cite{yadegari2015generic,saturn}. We tested this hypothesis and observe that this applies to the state-of-the-art tools \themida and \vmprotect, for both their fastest and strongest protection configurations: We found that a simple dead code elimination manages to reduce the number of assembly instructions per handler by at least 50\% for five different targets, tremendously simplifying both manual and automated analyses (\cf Table~\ref{tab:intro:commercial_vm_stats}). Subsequently, the resulting code can be further simplified using a wide range of automated techniques, including taint analysis~\cite{yadegari2014bit,yadegari2015generic}, symbolic execution~\cite{yadegari2015symbolic,yadegari2015generic}, program synthesis~\cite{blazytko2017syntia,david2020qsynth}, and various others~\cite{rolles2009unpacking,sharif2009automatic,coogan2011deobfuscation,kinder2012towards, barhelemy2016binary,eyrolles2016defeating,guinet2016arybo,bardin2017backward,eyrolles2017dissertation,liang2017deobfuscation,salwan2018symbolic,saturn}.

The reliance on \emph{syntactic} complexity in state-of-the-art obfuscation schemes and the broad arsenal of advanced deobfuscation techniques sparked further research in the construction of more resilient schemes that aim to impede these automated analyses.
Proposals were made to hinder taint analysis~\cite{cavallaro2007anti,sarwar2013effectiveness} and render symbolic execution ineffective~\cite{zhou2007information, banescu2016code, xu2018manufacturing, ollivier2019howto}. For example, the latter can be achieved by triggering a path explosion for the symbolic execution engine by artificially increasing the number of paths to analyze.
Other promising obfuscation schemes emerged, including Mixed Boolean-Arithmetic (MBA) expressions~\cite{zhou2007information,eyrolles2017dissertation,banescu2017predicting} that offer a model to encode arbitrary arithmetic formulas in a complex manner. The expressions are represented in a domain that does not easily lend itself to simplification, effectively hiding the actual semantic operations. Usually, automated approaches to deobfuscate MBAs are based on symbolic simplification~\cite{guinet2016arybo,barhelemy2016binary,eyrolles2016defeating,eyrolles2017dissertation}; they rely on certain assumptions about the expression's structure, making them unfit to simplify such expressions in the general case. Other approaches are based on program synthesis~\cite{blazytko2017syntia,david2020qsynth, menguy2021xyntia}, which have been proven highly effective for most tasks. In general, synthesis-based deobfuscation techniques remain unchallenged to date and are valuable methods for automated analysis of obfuscated code. Recent works aiming at simplifying MBA turned towards machine learning~\cite{feng2020neureduce} and algebraic simplification~\cite{liu2021mbablast}. Especially the latter approach, relying on a hidden two-way feature between 1-bit and n-bit variables used within MBAs, provides an automated attacker with unprecedented MBA deobfuscation capabilities.

\begin{table*}[htb]
\caption{VM handler statistics for \themida, \vmprotect, and our approach called \loki{}. The two commercial obfuscators are configured in their fastest (\emph{Virtualization} and \emph{Tiger White}) and strongest configuration (\emph{Ultra} and \emph{Dolphin Black}), but without additional security features (\eg anti-debug). We track their handlers' \emph{average number} of assembly and intermediate language (IL) instructions before and after dead code elimination (denoted as the percentage-wise reduction in parentheses). All values are averaged over five cryptographic algorithms (AES, DES, MD5, RC4, and SHA-1).}

\resizebox{\linewidth}{!}{
\begin{tabularx}{1.16\linewidth}{lRRRRR}
\toprule
 & \multicolumn{2}{c}{\vmprotect} & \multicolumn{2}{c}{\themida} & \multicolumn{1}{c}{\loki} \\
Statistics                                    & Virtualization & Ultra & Tiger White & Dolphin Black &\\
\midrule
Assembly instructions &     69 ($-50.79\%$) &   73 ($-51.58\%$) &  219 ($-53.68\%$) & 243 ($-56.01\%$) & 222 ($-1.14\%$) \\
IL instructions &     75 ($-50.88\%$) &   80 ($-51.89\%$) &  221 ($-53.76\%$) &   247 ($-55.94\%$) & 234 ($-1.44\%$) \\
\midrule
Handlers executed               &  46,591 & 151,303 & 83,191 & 290,815 & 4,123 \\
\dots of them unique              &  274 & 4,578 & 204 & 337 & 55 \\
\bottomrule
\end{tabularx}
}
\centering
\label{tab:intro:commercial_vm_stats}
\vspace{-1.0em}
\end{table*}

In this paper, we introduce a novel and comprehensive set of obfuscation techniques that can be combined to protect code against all known automated deobfuscation attacks, while imposing only reasonable overhead in terms of space and runtime.
Our techniques are specifically designed such that a human analyst gains no significant advantage from employing automated deobfuscation techniques, including compiler optimizations (\cf Table~\ref{tab:intro:commercial_vm_stats}), forward taint analysis, symbolic execution, and even program synthesis (\cf Section~\ref{sec:evaluation}). We explicitly assume scenarios where these techniques are specifically tailored to our design \emph{(white-box scenario)}.

To achieve such protection, we propose a generic algorithm to synthesize formally verified, arbitrarily complex MBA expressions. This is in strong contrast to state-of-the-art approaches that rely on a few handwritten rules, greatly limiting their effectiveness. For example, given $7,000$ VM handlers, \tigress---the state-of-the-art academic obfuscator---uses only 16 unique MBAs, while our design features \textasciitilde$5,500$ unique MBAs. As a result, our MBAs are highly unlikely to be simplified: In fact, current state-of-the-art MBA deobfuscation tools such as \mbablast~\cite{liu2021mbablast} can only simplify 0.5\% of \loki's MBAs.
Furthermore, we conduct the first conclusive analysis of the limits of program synthesis with regard to deobfuscation. Based on the resulting insights, we present a hardening technique capable of impeding program synthesis, reducing its success rate to 19\%---for \tigress, it is 67\%.
In summary, we present a new design featuring both high diversity and resilience against static and dynamic, automated deobfuscation attacks. While providing more value, our design incurs significantly less overhead compared to commercial, state-of-the-art obfuscation schemes (up to 40 times).
Moreover, we port modern testing techniques, \ie formal verification and fuzzing, to our design and show that complex combinations of obfuscation transformations benefit from such methods to assert the correctness of complex and non-deterministic obfuscation transformations.

\smallskip
\noindent
\textbf{Contributions.} We make the following contributions:
 \begin{itemize}
\item We present the design, implementation, and evaluation of \loki, a software obfuscation approach resilient against all known automated deobfuscation attacks, even in white-box scenarios. %
\item We introduce a generic approach to synthesize diverse and formally verified Mixed Boolean-Arithmetic (MBA) expressions of arbitrary complexity that withstand even current state-of-the-art deobfuscation attacks.
\item We are the first to propose an approach resilient against program synthesis-based attacks and map out limits of program synthesis in an empirical evaluation.
 \end{itemize}

\noindent We publish the source code of \loki as well as all evaluation artifacts (including test cases, binaries, and evaluation tooling) at \url{https://github.com/RUB-SysSec/loki}.

\section{Technical Background}\label{sec:background}

We start by providing an overview of the required technical information on obfuscation and deobfuscation techniques. 

\subsection{VM-based Obfuscation}\label{sec:background:VM}
Virtual machine-based obfuscation, also known as \emph{virtualization}, %
protects code by translating it into an intermediate representation called \emph{bytecode}. This bytecode is interpreted by a CPU implemented in software, adhering to a custom instruction set architecture (ISA). An attacker must first reverse engineer this software CPU, a tedious and time-consuming task~\cite{rolles2009unpacking,sharif2009automatic}. Only after understanding the VM, they can reconstruct the original high-level code. 

\textbf{VM Interpreter.}
The original, unprotected code is replaced with a call to the \emph{VM entry} that invokes the interpreter. It sets up the initial context of the VM and points it to the bytecode that is to be interpreted. This is implemented by the \emph{VM dispatcher} using a fetch-decode-execute loop: first, it fetches the next instruction, decodes its opcode, and then transfers execution to the respective \emph{VM handler}. Often, the handler is determined via a \emph{global handler table} that is indexed by the opcode. After handler execution, the control flow returns to the VM dispatcher. Eventually, execution finishes by invoking a special \emph{VM exit} handler aborting the loop.

\textbf{Abstraction of Handler Semantics.}\label{paragraph:abstraction_handler_semantics}
Handlers are often semantically simple~\cite{blazytko2017syntia,rolles2009unpacking}; they perform a single arithmetic or logical operation on a number of operands, \eg $x \odot y$. We call the semantic function of a handler, \ie the underlying instruction it implements, its \emph{core semantics}. We can represent core semantics as a function $f(x, y)$, or more general as $f(x, y, c)$ where $c$ is a constant. %
To measure the \emph{syntactic complexity} of the core semantics, we compute the (syntactic) expression depth of $f$ as the sum of all variable occurrences and operators. In contrast, the \emph{semantic depth} refers to the syntactic depth of the syntactically shortest equivalent expression. Intuitively, it can be understood as the number of nodes in an Abstract Syntax Tree (AST). 

\begin{example}
We can represent a VM handler's core semantics $x+y$ as $f(x, y, c) := x + y$ with a syntactic depth of $3$.  A syntactically more complex function $g(x,y, c) := x + y - x + c - c$ has a syntactic depth of $9$ but a semantic depth of $1$, since $g$ can be simplified to $g(x,y,c) := y$.
\end{example}

\textbf{Superoperators.}\label{paragraph:superoperators}
Superoperators~\cite{superoperators} are an approach to make handlers semantically more complex. Intuitively, this is achieved by combining different instruction sequences from the unprotected code into a single VM handler. 
Usually, these sequences compute independent results such that this VM \enquote{superhandler} computes multiple, independent VM handlers in a single step. As a consequence, superoperators often have multiple input and output tuples.
Related to our function abstraction, we can say the function $f_s((x_0, y_0, c_0), \dots, (x_n, y_n, c_n))$ computes an output tuple $(o_0, \dots, o_n)$, where $x_i$, $y_i$, $c_i$ and $o_i$ represent the core semantics' inputs/output of a semantically simple VM handler. While originally developed to minimize the number of handlers executed to improve performance, superoperators have been used by obfuscators such as \tigress primarily for obtaining more complex VM handlers.

\subsection{Mixed Boolean-Arithmetic}
\emph{Mixed Boolean-Arithmetic (MBA)} describes an approach to encode expressions in a syntactically complex manner. The goal is to hide underlying semantics in syntactically complex constructs. First described by Zhou~\etal~\cite{zhou2007information}, MBA algebra connects arithmetic operations (\eg addition) with bitwise operations (\eg logical operations or bitshifts). The resulting expressions are usually hard to simplify symbolically~\cite{eyrolles2017dissertation,xu2021boosting}, since, for every expression, an infinite number of syntactic representations exists. In general, the task of reducing MBA expressions---known as \emph{arithmetic encodings}~\cite{banescu2016code}---to equivalent but simpler expressions is NP-hard~\cite{zhou2007information}.

\begin{example}
$f(x, y, c) := x + y$ and $g(x, y, c) := (x \oplus y) + 2 \cdot (x \land y)$ are semantically equivalent. 
Both implement the same core semantics, but $g$ uses a syntactically more complex representation, called MBA.
\end{example}

\section{Automated Deobfuscation Attacks}~\label{sec:background:attacks}

In the following, we detail common techniques used to analyze obfuscated code.

\textbf{Forward Taint Analysis.}
\emph{Forward taint analysis} follows the data flow of so-called \emph{taint sources}, \eg input variables, and marks all instructions as \emph{tainted} that directly or indirectly depend on these sources~\cite{schwartz2010all, yadegari2015generic, yadegari2014bit, sharif2009automatic}. Taint analyses are implemented with varying granularity, referring to the smallest unit they can taint. Common approaches use either bit-level or byte-level granularity.
Forward taint analysis can be used to reduce obfuscated code to the instructions depending on user input. The underlying idea is that important semantics rely only on the identified taint sources. All other code constructs, \eg as added by an obfuscator, can be omitted in an automated matter. Still, if these constructs perform calculations on the user input, taint analysis can be mislead~\cite{cavallaro2007anti,sarwar2013effectiveness}.
\begin{example}\label{example:taint}
In Figure~\ref{fig:codesnippet}, assume \texttt{eax} is a taint source. The analysis taints the first, third, and fourth instruction since they propagate a taint source. It does \emph{not} taint the second instruction. While its value is later used in tainted instructions, it does not directly depend on \texttt{eax}.
\end{example}

\textbf{Backward Slicing.}
Contrary to forward taint analysis, backward slicing is a backward analysis. Starting from some output variable, it recursively backtracks and marks all input variables on which the output depends~\cite{yadegari2015generic,weiser1981program,danicic2018static}. 
In code deobfuscation, slicing can be used to find all instructions that contribute to the output. Applied to VM handlers, it allows to strip all code not directly related to a handler's core semantics. Similar to forward taint analysis, increasing the number of dependencies (\eg by inserting junk calculations to the output) reduces the usefulness of slicing.
\begin{example}
When backtracking the value of \texttt{edx} (line 4 in Figure~\ref{fig:codesnippet}) by following each use and definition, each instruction is marked as they all contribute to the output.
\end{example}

\begin{figure}[tp]
\begin{minted}[numbersep=0.5em,fontsize=\small,frame=leftline,baselinestretch=0.9,linenos]{nasm}
    mov edx, eax         ; edx1 := eax1
    mov ecx, 0x20        ; ecx1 := 0x20
    add edx, ecx         ; edx2 := edx1 + ecx1
    add edx, 0x10        ; edx3 := edx2 + 0x10
\end{minted}
\caption{An assembly code snippet used to illustrate forward taint analysis, backward slicing, and symbolic execution.}%
\label{fig:codesnippet}%
\vspace{-1.5em}
\end{figure}

\textbf{Symbolic Execution.}
Symbolic execution allows to summarize assembly code algebraically. Instead of using concrete values, it tracks symbolic assignments of registers and memory in a state map~\cite{schwartz2010all}. Often, it works on a verbose representation of code, called \emph{intermediate language (IL)}.
Symbolic executors usually know common arithmetic identities and can perform basic simplification, \eg constant propagation. 
Applied to code obfuscation, symbolic execution is used to symbolically extract the core semantics of VM handlers~\cite{liang2017deobfuscation}, track user input in an execution trace~\cite{yadegari2015symbolic,yadegari2015generic,salwan2018symbolic}, or detect opaque predicates (in combination with SMT solvers)~\cite{bardin2017backward}. Typically, techniques to impede symbolic execution aim at artificially increasing the syntactic complexity of arithmetic operations (via MBAs) or the number of paths to analyze (triggering a so-called \emph{path explosion})~\cite{ollivier2019howto,banescu2016code}.
\begin{example}
After symbolic execution of Figure~\ref{fig:codesnippet}, we obtain the following mappings: \texttt{eax} maps to itself (it has not been modified), \texttt{ecx} maps to $\mathrm{0x20}$ (line 2). The  formula for \texttt{edx} is $\texttt{eax} + \mathrm{0x20} + \mathrm{0x10}$. Using arithmetic identities, the symbolic execution engine can simplify the expression to $\texttt{eax} + \mathrm{0x30}$.
\end{example}

\textbf{Program Synthesis.}\label{paragraph:program_synthesis}
In contrast to other techniques that rely on syntactic analysis of obfuscated code, \emph{program synthesis}-based approaches operate on the semantic level. They treat code as a black box and attempt to reconstruct the original code based on the observable behavior, often represented in the form of input-output samples. Approaches such as \syntia~\cite{blazytko2017syntia} and \xyntia~\cite{menguy2021xyntia} attempt to find an expression with equivalent behavior by relying on a stochastic algorithm traversing a large search space. Other approaches, \eg \qsynth~\cite{david2020qsynth}, are based on enumerative synthesis: they compute large lookup tables of expressions which they use to simplify parts of an expression, reducing its overall complexity.
For code deobfuscation, these approaches are used to simplify syntactically complex constructs (\eg MBAs) or to learn  semantics of VM handlers. 
However, program synthesis struggles with finding semantically complex expressions. 
\begin{example}
Consider the function $f(x,y, c) := (x \oplus y) + 2 \cdot (x \land y)$. To learn $f$'s core semantics, we generate
random inputs and observe $f(2, 2, 2) = 4$, $f(10, 13, 10) = 23$, and $f(16, 3, 0) = 19$. A synthesizer eventually produces a function $g(x, y, c) :=  x + y$ that has the same input-output behavior. Notably, it learns that parameter $c$ is irrelevant.
\end{example}

Ideally, superoperators provide such expressions. However, our experiments (\cf Section~\ref{sec:evaluation:semantic_attacks}) demonstrate that current designs (\eg as used by \tigress) are still vulnerable; since superoperators combine different core semantics (represented as individual inputs/output tuples), an attacker can synthesize each core semantics separately by targeting each output $o_i$.

\textbf{Semantic Codebook Checks.}
A semantic codebook contains a list of expressions that an attacker expects to exist within obfuscated code. For a syntactically complex expression $f$, an attacker checks if $f$ is semantically equivalent to an expression $g$ in the codebook by using an SMT solver~\cite{vanegue2012smt}. If the SMT solver cannot find an input \emph{distinguishing} $f$ and $g$, it \emph{formally proved} they behave the same for all possible inputs.
A typical application scenario are VM handlers: They often implement a simple core semantics (\eg $x + y$)~\cite{rolles2009unpacking,blazytko2017syntia}. Thus, an attacker can construct a codebook based on simple arithmetic and logical operations. 
As codebooks must contain the respective semantics, increasing the semantic complexity of expressions requires an (exponentially) larger codebook, making the approach infeasible for a practical application.
\begin{example}
Consider a function  $f(x, y, c) := (x \oplus y) + 2 \cdot (x \land y)$ and a codebook $\mathrm{CB}: =\{x - y, x \cdot y, x + y, \dots\}$. An attacker can consecutively pick an entry $g(x,y, c)\in \mathrm{CB}$ and verify whether $f = g$ using an SMT solver. To this end, the solver searches an assignment that satisfies $f(x, y, c) \neq g(x,y, c)$. Only for $g(x,y, c) := x + y$ no solution can be found. Thus, the attacker proved that $f$ can be reduced to a syntactically shorter expression, $x + y$.
\end{example}

\section{Design}\label{sec:design}

We envision a combination of obfuscation techniques where the individual techniques harmonize and complement each other to thwart automated deobfuscation attacks. 
In line with this philosophy, we now present a set of generic techniques where each constitutes a defense in a particular domain. However, when these techniques are effectively combined, they exhibit comprehensive protection against automated attacks. %
To achieve lasting resilience, we focus on inherent weaknesses underlying existing automated attack methods, instead of targeting specific shortcomings of a given implementation. We further underline our techniques' generic nature by discussing their application on an abstract function $f(x, y, c)$ as introduced in Section~\ref{paragraph:abstraction_handler_semantics}.
Next, we first discuss the design principles of our approach, present the attacker model, and afterwards explain the individual techniques in detail.

\subsection{Design Principles}

We have seen outlined automated attack methods can succeed in extracting a function $f$'s core semantics (\cf Section~\ref{sec:background:attacks}).
To mitigate these attacks, our design is based on four principles:
\begin{inparaenum}[(1)]
\item merging core semantics,
\item diversifying the selection mechanism,
\item adding syntactic complexity, and
\item adding semantic complexity.
\end{inparaenum}
In the following, we present techniques incorporating these principles and discuss their purpose as well as synergy effects emerging for our overall design.

\textbf{Merging Core Semantics.}
Our first technique extends $f$ by merging different, independent core semantics to increase the complexity. This can be understood as combining different, independent VM handlers in a single handler, or---in a more generic setting---combining different semantic operations of an unprotected unit of code in a single function $f$.
The merge is facilitated in such a way that each core semantics is always executed. Still, as these semantics are independent of each other, we must ensure they are individually addressable, \ie $f$'s output is equivalent to the result of a specific core semantics.
To allow the selection of the desired semantics, we extend the function definition to $f(x, y, c, k)$, where $k$ is a \emph{key} selecting the targeted core semantics.
Formally, the selection is realized by introducing a point function $e_i(k)$, called \emph{key encoding}, that is associated with a specific core semantics and returns $1$ only for its associated key, $0$ for other valid keys. Essentially, the $e_i(k)$ are \emph{(partial) point functions} guaranteeing that only the selected semantics' output is propagated despite executing all semantics.
A consequence of this interlocked, \enquote{always-execute} nature is that taint analysis and backward slicing fail to remove all semantics in $f$ not associated with a specific $k$. %

\begin{example}
We want to design a function $f$ that returns, based on a distinguishing key, either $x + y$ or $x - y$. We write this as $f(x, y, c, k) := e_0(k)(x + y) + e_1(k)(x - y)$ where $e_i(k)$ can be any point function returning $1$ for the associated $k$ and $0$ otherwise, for example $e_0(k) := k == 0xdead$. Assuming that $e_0(k)$ returns $1$ and $e_1(k)$ yields $0$, $f$ returns $x + y$.
\end{example}

\textbf{Diversifying Selection.}
Assuming a function $f$ contains multiple merged semantics, an attacker's goal might be to extract operations by comparing them to a semantic codebook. For static attackers (\cf Section~\ref{sec:threat_model}) using an SMT solver, this implies finding a key satisfying the core semantics' associated key check.
Therefore, we propose to arithmetically encode the key checks such that SMT solvers struggle with finding a satisfying key. However, relying on a specific encoding (\eg based on factorization), may allow an attacker to identify patterns and adapt their attack to the specific problem at hand, e.g., by using domain-specific solutions such as brute-forcing the factors.
Accounting for this attack vector, we propose a second type of key encoding, which relies on the synthesis of partial point functions tailored to the respective key check. The synthesis creates a diverse and unpredictable set of functions that hinder efficient pattern matching.
Using both key encodings within a function $f$, we ensure that SMT solvers struggle while maintaining sufficient diversity to render specialized attacks inefficient.

\textbf{Adding Syntactic Complexity.}
Assuming merged semantics using different key encodings, an attacker can still differentiate between key encoding and core semantics for a given function $f$, as $e_i(k)$ operates only on the key while the core semantics use $x$, $y$, and $c$. At the same time, a dynamic attacker with knowledge of $k$ can employ symbolic execution to simplify $f$ to the core semantics associated with the known $k$ by arithmetically nullifying operations not contributing to the result. To prevent such an attack, we increase the syntactic complexity by adding Mixed Boolean-Arithmetic (MBA) formulas to key encodings as well as core semantics.

Symbolically executing these syntactically complex formulas creates no meaningful expressions. Even though modern symbolic execution engines feature simplification rules for basic arithmetic identities and laws, there exists an unlimited number of MBA representations. In general, simplifying such an expression to its syntactically smallest representative is NP-hard~\cite{zhou2007information}.

\begin{example}
For $f(x, y, c ,k) := e_0(k)(x+y) + e_1(k)(x-y)$, we can replace $x + y$ with $(x \oplus y) + 2 \cdot (x \land y)$, $x-y$ with $x + \neg y + 1$, and replace the multiplication of $e_1(k)(x-y)$ with the rule $(a\land b)\cdot(a\lor b) + (a\land\neg b)\cdot(\neg a\land b)$ for $a\cdot b$, resulting in the final function $f(x, y, c ,k) := e_0(k)((x \oplus y) + 2 \cdot (x \land y)) + (e_1(k)\land (x + \neg y + 1))\cdot(e_1(k)\lor (x + \neg y + 1)) + (e_1(k)\land\neg (x + \neg y + 1))\cdot(\neg e_1(k) \land(x + \neg y + 1))$.
\end{example}

To exploit this weakness of symbolic execution and provide a high diversity, we \emph{synthesize} and \emph{formally verify} MBAs instead of using hardcoded rules. This additionally complicates pattern matching and increases the number of instructions marked by forward taint analysis and backward slicing.

\textbf{Adding Semantic Complexity.}
One of the remaining problems are semantic attacks, for example, a dynamic attacker that performs a codebook check or uses input-output behavior to learn an expression equivalent to the core semantics (\eg via program synthesis). Therefore, we increase the core semantics' complexity by applying the concept of superoperators (\cf Section~\ref{sec:background:VM}). These superoperators make core semantics arbitrarily long and increase the search space for semantic attacks drastically.

\begin{example}
Instead of using core semantics of depth 3 (\eg $x + y$), we apply more advanced core semantics such as $(x + y) \cdot (x\oplus y))$ with depth 7 or $((x \cdot c) \ll (y \lor (x \oplus c)))$ with depth 9, resulting in $f(x ,y ,c, k)$ := $e_0(k)((x + y) \cdot (x \oplus y)) + e_1(k)((x \cdot c) \ll (y\,|\,(x \oplus c)))$.
\end{example}
While superoperators increase the semantic and syntactic complexity of core semantics, we further extend their syntactic complexity using MBAs. Their synergy additionally diminishes the effect of automated attacks.

\subsection{Attacker Model}\label{sec:threat_model}
Intuitively, we envision a strong attacker to measure how our obfuscation scheme fares under worst-case conditions. For this purpose, we assume that an attacker has access to all automated attacks (\cf~Section~\ref{sec:background:attacks}).

We assume an attacker has access to the target binary that includes at least one well-defined unit of obfuscated code at a known location. 
In line with our previous abstraction, we say this code unit can be represented by a function $f(x, y, c, k)$. 
The attacker's goal is to reconstruct the core semantics of $f$ associated with a specific $k$.  We require the reconstructed semantics to
\begin{inparaenum}[(1)]
\item contain \emph{only} the core semantics associated with the specified $k$ and %
\item be comparable to the original code's semantics in terms of syntactic complexity. %
\end{inparaenum}%
The intuition behind these constraints is to exclude trivial solutions such as providing the unmodified function $f$ itself (which contains, amongst others, the core semantics for the required $k$).

Further, we assume two \emph{types} of attackers, a static and a dynamic one. The \emph{static attacker} knows the precise code locations of $x$, $y$, $c$, and $k$ as well as the location of function $f$'s output. As a result, they can enrich static analyses, \eg by defining these code locations as taint sources. 
A \emph{dynamic attacker} extends the former by the ability to inspect and modify the values at these code locations. 
In particular, they can observe any key $k$ and propagate it to remove core semantics not associated with this~$k$.
While a dynamic attacker is more powerful (in terms of accessible information), certain analysis scenarios such as code running on specific hardware (\eg embedded devices), analysis on function-level without context, or the presence of techniques like anti-debugging~\cite{antidebugging,antidebugging2} may rule out dynamic analysis in practice.

\subsection{Key Selection Diversification}\label{sec:design:key_selection_diversification}
We want to prevent static attackers from learning the core semantics via semantic codebook checks and prevent identification of patterns in the key selection. To do so, we employ two different key encoding schemes: Key selection based on (1) the factorization problem, and (2) synthesized partial point functions.
To conduct a semantic codebook check, an attacker uses an SMT solver to check for each entry of the codebook whether it is semantically equivalent to $f$.
Assuming that $f$ indeed includes a matching core semantics, the SMT solver has to find a value for $k$ such that the corresponding $e_i(k)$ evaluates to $1$.
One way to prevent this is to design a key encoding that relies on inherently hard problems for the SMT solver, such as factorization.

\textbf{Factorization-based Key Encoding.}
Factorization of a semi\-prime $n$ (the product of two primes, $p$ and $q$) is an inherently hard problem \emph{as long as} the size of the factors are large enough (commonly, a few thousand bits).
SMT solvers prune the search space by learning partial solutions for a given problem~\cite{kroening2008decision}, but since no partial solutions exist for factorization, they are forced to perform an exhaustive search.

We define our factorization-based key encoding as  $e_i(k):=(n\ \operatorname{mod}\ k) \equiv 0$ where $k$ is a valid 32-bit integer representing one of the two factors ($k \not\in \{1,n\}$).
As our evaluation shows, this encoding suffices to stall SMT solvers. 
However, its distinct structure makes pattern matching attempts easy. To increase diversity, we use MBAs and a second key encoding.

\textbf{Partial Point Functions.}
Instead of restraining our set of key encodings to a specific type, we synthesize generic point functions without any predefined structure. 
This is based on the insight that the $e_i(k)$ impose only a single constraint: they must be defined for all valid keys (returning $1$ for their associated one, $0$ for others). Invalid keys may return arbitrary values, making our synthesized functions \emph{partial point functions}. 
Consequently, we are not restricted to specific point functions, such as the factorization-based encoding, but can use arbitrary point functions fulfilling this constraint. 

Given a grammar containing ten different arithmetic and logical operations (such as addition, multiplication, and logical and bitwise operations), we generate expressions by chaining a randomly selected operation with random operands. This operand is either an arbitrary key byte or a random 64-bit constant. We chain at most 15 operations to limit the overhead resulting from this expression. Finally, we check if the synthesized expression satisfies the point function's constraint.
\begin{example}
Let  $(k_0, k_1, k_2) := (\mathrm{0x1336},~\mathrm{0xabcd},~\mathrm{0x11cd})$ be a set of keys. Then, we synthesize the point function $e_0(k) :=((\mathrm{0xff} \land k) \oplus \mathrm{0xcd}) \cdot \mathrm{0x28cbfbeb9a020a33}$ for a 64-bit vector $k$. $e_0(k)$ evaluates to $1$ for $k_0$ and to $0$ for $k_1$ and $k_2$. 
For all other keys, it returns arbitrary values.
\end{example}

\subsection{Syntactic Complexity: MBA Synthesis}\label{sec:design:mba_synthesis}

To thwart symbolic execution and pattern matching, we use MBAs for all components, including core semantics and key encodings.
As hardcoded rules only provide low diversity, we precompute large classes of semantically equivalent arithmetic expressions and combine them through recursive, randomized expression rewriting. We now detail the creation of the equivalence classes and discuss our term rewriting.

\textbf{Equivalence Class Synthesis.}
To create semantic equivalence classes for expressions, we rely on enumerative program synthesis~\cite{gulwani2010dimensions,bansal2006automatic}. To this end, we first define a context-free grammar with a single non-terminal symbol $S$ as start symbol and two terminal symbols, $x$ and $y$, representing variables. For each arithmetic operation, we define a production rule that maps the non-terminal symbol to arithmetic operations (\eg addition) or terminal symbols. To apply a specified production rule to a non-terminal expression, we replace the left-most $S$ with the rule. Expressions without a non-terminal symbol can be evaluated by assigning concrete values to $x$ and $y$. We say that the \emph{depth} of an expression represents the number of times a non-terminal symbol was replaced by a production rule.

\begin{example}
The grammar $(\{S\},\Sigma = V \cup O, P, S)$ with the variables $V=\{x, y\}$, the set of arithmetic symbols $O = \{+, -\}$ and the production rules
$P=\{S\rightarrow x \; | \; y \; | \;  (S \; + \; S)  \; | \;  (S \; - \; S)\}  $ defines the syntax of how to generate terminal expressions. To derive the expression $x + y$ of depth $3$, we apply the following rules: $S\rightarrow (S + S) \rightarrow (x + S)\rightarrow (x + y)$. With a mapping of $\{x \mapsto 2, y \mapsto 6\}$, we can evaluate the terminal expression to $8$.
\end{example}

We now describe how we use our context-free grammar in combination with Algorithm~\ref{algo:equivalence_classes}, which illustrates the high-level approach of equivalence class synthesis.
Starting with a worklist of non-terminal states (initialized with the start symbol $S$), we iteratively process all expressions for a certain depth until we reach a specified upper bound depth $N$. For a given depth, we derive all terminal and non-terminal expressions (also referred to as \emph{states}) before processing the terminals and then repeating the process for the next depth. The call to \texttt{process\_terminals} is responsible for sorting the expressions into the respective equivalence classes. To this end, we evaluate all expressions for a high number of different inputs (\eg $1,000$), recording their output. Expressions with the same output behavior for all provided inputs are sorted into the same equivalence class.
This provides an effective but coarse-grained sorting of expressions into potential equivalence classes. In a final step, we verify that these classes are semantically correct. For this, we choose the member with the smallest depth as representative and check with an SMT solver that all other members are semantically equivalent to this representative. Expressions failing this check are removed from the equivalence class. All remaining expressions are formally proven to not alter the original semantics.

\begin{algorithm}[tp]
\small
  \KwData{\inlinecode{n} is the maximum depth.
  }
\inlinecode{states} $\leftarrow$ $\{S\}$

\For{$d\leftarrow 1$ \KwTo $n$}{

\inlinecode{terminals} $\leftarrow$ \inlinecode{derive\_terminals(states)}

\inlinecode{process\_terminals(terminals)}

\inlinecode{non\_terminals} $\leftarrow$ \inlinecode{derive\_non\_terminals(states)}

\inlinecode{states} $\leftarrow$ \inlinecode{non\_terminals}
}
\caption{\small Computing equivalence classes.}%
\label{algo:equivalence_classes}%
\vspace{-0.6em}
\end{algorithm}%

To prune the search space and avoid trivial expressions (\eg $x + 0$), we symbolically simplify each terminal and non-terminal expression. For this purpose, we apply a normalization step to commutative operators, 
perform constant propagation, and simplify based on common arithmetic identities (\eg $x + y - y$ becomes $x$).

\textbf{Expression Rewriting.}
So far, we generated a large set of diverse equivalence classes we can use for replacing syntactically simple expressions with more complex ones.  A naive approach replacing expressions with MBAs from the equivalence classes is bounded by the largest depth found in the respective class. To overcome this limitation, we propose a recursive expression rewriting approach using the equivalence classes as building blocks. This allows us to create expressions of \emph{arbitrary} syntactical depth. 
Even assuming an attacker is in possession of all rewriting rules, it is difficult to invert an expression: Term rewriting is inherently \emph{destructive}~\cite{willsey2021egg}. Without knowing the applied rewriting rules and their order, an attacker has to check all possibilities: $n$ rewriting rules applied over $m$ layers, resulting in the prohibitively large number of $n^m$ candidates.

Given some expression $e$, we pick a random subexpression and check if it is a member of an equivalence class. If it is, we randomly choose another member from this class and use it to replace the picked subexpression within $e$. We recursively repeat this process for a randomly determined upper bound $n$.
As all members within an equivalence class are proven to be pairwise equivalent, each replacement is guaranteed to produce an equivalent expression. Consequently, the final expression is provably equivalent to the first.

\begin{example}
Assume that we want to increase the syntactic complexity of $e := (x + y) + z$  with the upper bound $n=2$. First, we randomly choose the subexpression $x + y$. We then pick another member of the same equivalence class---$(x \oplus y) + 2 \cdot (x \land y)$---and replace it in $e$. In this case, we obtain $e := ((x \oplus y) + 2 \cdot (x \land y)) + z$. In a second step, we choose $x\oplus y$, pick the semantically equivalent member $(x \lor y) - (x \land y)$ and replace it again. The final MBA-obfuscated expression is $e := (((x \lor y) - (x \land y)) + 2 \cdot (x \land y)) + z$.
\end{example}

Empirical testing showed 
that for an initial expression the randomly picked subexpressions would often be short, causing the resulting recursive rewriting to be very local in nature rather than considering all of the expression. The previous example illustrates this behavior. Considering the expression as an \emph{abstract syntax tree (AST)}, we twice replaced deeper parts of the AST while ignoring the top-level operation (addition with $z$). Consequently, subsequent iterations would be even less likely to pick the high-level operation, considering the wealth of other operations to pick from. Therefore, the AST would be significantly unbalanced. To avoid this, we prefer selecting top-level operations in the first loop iterations.

\subsection{Semantic Complexity: Superoperators}\label{sec:design:superoperators}
Up to this point, $f$'s core semantics have a rather low semantic complexity (\eg $x + y$). To thwart semantic attacks, we use a variation of superoperators that increase the semantic complexity. 
The intention is to significantly increase the search space for an attacker:
For example, assume a set of three variables $V$ and a set of six binary operations $O$: For semantic depth 3 (\eg $x + y$), an expression contains $m = 2$ variables and $n = 1$ operations, such that an attacker has to brute-force at most $|V|^{m} * |O|^{n} = 3^2 * 6^1 = 54$ possibilities. For depth $7$ (\eg $((x + y) \cdot (x \oplus c))$), they must try up to $3^4 +  6^3 = 17,496$ different expressions (or $314,928$ for depth 9).
In other words, the search space grows exponentially, making semantic code book checks as well as program synthesis infeasible.

However, common superoperator strategies, \eg as used by Tigress~\cite{tigress}, are not resilient against these attacks (\cf Section~\ref{sec:evaluation:semantic_attacks}).
They usually include independent core semantics, each having their own output; this causes the handler to have multiple, independent outputs, which an attacker can target individually.
As each core semantics itself usually implements only a single operation~\cite{rolles2009unpacking,blazytko2017syntia} (\eg $x + y$ with semantic depth 3) attacking one such superoperator is similar to attacking a series of regular handlers.
To avoid this pitfall, we design our superoperators to preserve the signature of $f$ (a \emph{single} output and $x$, $y$, $c$ and $k$ as inputs) while providing a high semantic depth. 
In other words, our superoperators consist of a chain of core semantics that depend on each other and must be executed sequentially: The output of the core semantics is used as input for subsequent core semantics; the last core semantics produces the output of the handler.
Even if an attacker is aware of these superoperators, they cannot split a handler into multiple separate synthesis tasks and forces them to synthesize the whole expression.

On a technical level, we construct superoperators based on data-flow dependencies, more precisely \textit{use-definition chains} based on \emph{static single assignment (SSA)}~\cite{cytron1989ssa}: 
Given an unprotected code unit in form of instructions in three-address code, we assign a unique variable to each variable definition and replace subsequent variable uses with its latest definition on the right-hand side (called \emph{SSA form}). Then, we build superoperators by first randomly picking variables on the right-hand side and then replacing these \emph{uses} by their respective variable \emph{definitions} recursively. By choosing lower and upper limits for the recursion bound, we can control the superoperators' semantic depth. To further increase the syntactic complexity, we apply our MBAs.

\begin{example}
Assume we have three sequential instructions (Figure~\ref{fig:sueroperator_snippet}, l. 1-3) implementing semantically simple operations; each represents an individual core semantics. 
Notably, the first instruction's output serves as input for the second and third. Similarly, the second instruction is an input to the third.
To create a superoperator that implements a semantically more complex operation, we transform the code into SSA form, (randomly) pick \texttt{b1} in the third instruction and replace this use by its definition (l. 2), yielding \texttt{d2 := (a * d1) | d1}.  
When picking \texttt{d1}, we replace it by its definition (l. 1) accordingly, transforming the expression into \texttt{d2 := (a * (a + b)) | (a + b)}. While the initial expressions have a semantic depth of 3, the superoperator's depth is 9.
\end{example}

\begin{figure}[tp]
\begin{minted}[numbersep=0.5em,fontsize=\small,frame=leftline,baselinestretch=0.9,linenos]{bash}
    d := a + b          ;  d1 := a  + b
    b := a * d          ;  b1 := a  * d1
    d := b | d          ;  d2 := b1 | d1
\end{minted}
\caption{Three different core semantics, each implementing a simple operation. On the right-hand side, the SSA form of the respective expressions.}%
\label{fig:sueroperator_snippet}%
\vspace{-1.5em}
\end{figure}

Intuitively, replacing a \emph{use} by its respective \emph{definition} is guaranteed to preserve the semantics, as variable assignments are immutable in SSA form. 
Additionally, we prove the rewritten superoperator is equivalent to the original code with symbolic execution.

\subsection{Synergy Effects}
To summarize, each of our components thwarts specific deobfuscation attacks: MBAs tackle symbolic execution and pattern matching, while the nature of $f$ with its multiple core semantics, selected via a key, prevents taint analysis and backward slicing from removing irrelevant semantics. Further, our key encodings render semantic codebook checks infeasible superoperators increase the semantic complexity, throwing off semantic attacks.

As indicated, especially the combination of our techniques prevents automated deobfuscation attacks: They do not only co-exist but have beneficial synergy effects, which in turn improve the overall resilience of the combination. For example, our MBAs weaken pattern matching on all levels, including key encodings, and cause the differences between key encoding and core semantics to blur. Besides the syntactic confusion introduced, we can propagate the core semantics into the key encoding and vice versa. For instance, we may use MBAs that extend the key check with the variables $x$ or $y$ using arithmetic identities that do not alter the key check itself.
At the same time, MBAs benefit from superoperators given they provide ample opportunity to pick and replace subexpressions.

\subsection{Verification of Code Transformations}

Obfuscation generally modifies the syntactic representation of code; thus, it is crucial to verify that it does not change the code's semantic behavior. One can achieve this by checking if the transformed code is semantically equivalent to the original one. While this works well for short sequences of instructions (\eg by using SMT solvers) within a reasonable amount of time, it does not scale to complex programs. In such cases, the industrial state of the art approximates these guarantees by using extensive random testing~\cite{afl,klees2018evalfuzz}.%

For our design, we choose the best applicable verification method to ensure correctness: For individual components, we use formal verification to prove their correctness (\cf Sections~\ref{sec:design:key_selection_diversification},~\ref{sec:design:mba_synthesis},~\ref{sec:design:superoperators}).
To improve the confidence of the correctness of the combination, we use an approach similar to black-box fuzzing~\cite{radamsa,peach}, where we compare the I/O behavior of the original and transformed code for a user-configurable number of random inputs, usually ranging from $1,000$ to $10,000$. These are randomly sampled depending on the type expected by the program (\eg ASCII strings, random 64-bit integers, or known edge cases such as $0$ or $0xff..ff$), which needs to be specified by the user. Crucially, we rely on human insight and careful specification of the input domain such that the sampled inputs cover the full program functionality. We apply this fuzzing both on the binary level as well as on the intermediate representation; for the former, we compare the compiled versions of the unprotected and protected programs, while we emulate the program's intermediate representation before and after transformations for the latter. 
As a consequence of our handlers' interlocked, always-execute nature, we achieve full code coverage and path coverage both on the intermediate representation as well as on the binary level for all handlers needed to represent the original code.

\section{Implementation}\label{sec:implementation}

To evaluate our techniques, we implement a VM-based obfuscation scheme named \loki on top of \llvm~\cite{llvm_homepage} (version 9.0.0) and a code transformation component written in Rust. \loki consists of \textasciitilde3,100~LOC in C++ and \textasciitilde8,700~LOC in Rust. In this scheme, each function $f(x, y, c, k)$ is represented by one of our 510 handlers. In other words, each handler can implement any semantic operation that requires no more than two input variables and one constant. Our handlers support the inclusion of three to five core semantics (randomly chosen at creation time), which can be addressed by setting $k$ accordingly. Besides these 510 handlers, we have a \emph{VM exit} and a handler managing memory operations. The control flow between handlers is realized as \emph{direct threaded code}~\cite{klint1981interpretation}, \ie each handler inlines the VM dispatcher. Our VM assumes a 64-bit architecture. Code operating on smaller bit sizes is semantically upcasted to guarantee correctness.

Our approach to obfuscate real-world code consists of three major steps: Lifting, code transformation, and compilation. The \emph{lifting} starts with a given C/C++ input program that contains a specified function to protect. 
We then translate this function to \llvm's intermediate representation (IR) and use various compiler passes to optimize the input and unroll loops as our prototype does not support control-flow to reduce engineering burden. Note that this is no inherent limitation of our approach, but a simplification we made as \llvm's passes sufficed in creating binaries that our prototype implementation can process.
Finally, we lift the resulting LLVM IR to a custom IR which the code transformation component internally operates on. This component \begin{inparaenum}[(a)]
\item parses the lifted representation of the targeted function,
\item creates superoperators based on this input (with recursion bound $3$ to $12$),
\item instantiates the VM handlers, applies our obfuscation techniques (\eg MBAs), and verifies them. 
For MBAs, we use a random recursive expression rewriting bound between $20$ and $30$. We choose from a pre-computed database of $843,467$ MBAs (all expressions up to a depth of $9$), split over $48$ equivalence classes. In each class, there are roughly 17,500 entries on average.
To exemplify the dimensions: An attacker has to try up to $n_{Loki}^m = 843,467^{30} = 6.1*10^{177}$ possibilities to simplify our MBAs; Based on our reverse engineering efforts, state-of-the-art obfuscator \tigress features only $47$ hand-crafted rules (that are not applied recursively), such that an attacker has to evaluate $n_{Tig}^{m} = 47^1 = 47$ possibilities.
\item Finally, the Rust component generates the VM bytecode and translates the handlers back into \llvm IR%
\end{inparaenum}.
Then, obfuscated and original code are compiled with \texttt{-O3} and verified.

\section{Experimental Evaluation}\label{sec:evaluation}

Based on our prototype implementation, we evaluate if our approach can withstand automated deobfuscation techniques (\emph{resilience}), while maintaining \emph{correctness} and imposing only acceptable overhead (\emph{execution cost}). Overall, we follow the evaluation principles outlined by Collberg \etal~\cite{collberg1998manufacturing}.

All experiments were performed using Intel Xeon Gold 6230R CPUs at 2.10 GHz with 52 cores and 188 GiB RAM, running Ubuntu. %
Our obfuscation tooling uses LLVM~\cite{llvm_homepage} (v. 9.0) and the SMT solver \zt~\cite{z3-homepage} (v. 4.8.7). For tracing coverage, we rely on Intel Pin~\cite{intelpin} (v. 3.23). Our deobfuscation tooling is based on \miasm~\cite{miasm-github} (commit \texttt{65ab7b8}), %
\triton~\cite{salwan2015triton} (v. 0.8.1), and \syntia~\cite{blazytko2017syntia} (commit \texttt{e26d9f5}).
We use our prototype of \loki and the academic state-of-the-art obfuscator, \tigress~\cite{tigress} (v. 3.1), to obfuscate five different programs, each implementing a cryptographic algorithm: AES, DES, MD5, RC4, and SHA1. 
This is a common approach: the first three are based on an obfuscation data set provided by Ollivier~\etal~\cite{ollivier2019howto}; the others are adapted from reference implementations~\cite{rc4implementation,sha1implementation}. 
These algorithms are representative for real-world scenarios in which cryptographic algorithms are used to guard intellectual property (\eg hash functions used for checksums in commercial DRM systems)~\cite{nakanishi2020intertwining}. 
In a case study, we obfuscate \vlc's DVD decryption routine to show how \loki can be applied onto real-world use cases.
Where necessary, we adapt the programs slightly to allow \loki to process them (\cf Section~\ref{sec:implementation}) without modifying their functionality. 
\tigress' configuration (\cf Appendix~\ref{sec:appendix:tigress}) resembles our design and works on the same source code files.

\subsection{Benchmarking}\label{sec:benchmarking}

Our goal is to benchmark the \emph{correctness} and \emph{cost} of our obfuscator. We do so by conducting a series of experiments, measuring the overhead in terms of runtime and disk size as well as verifying the correctness of transformed code. 
For each obfuscator, we create $1,000$ obfuscated instances for each of the five targeted programs and use them for all experiments. The overhead comparison is given as factor relative to the original, unobfuscated program compiled with \texttt{-O3}. To measure the MBA overhead, we create another $1,000$ obfuscated instances without any MBAs for \loki.

\begin{experiment}[Correctness]\label{experiment:correctness}
For each target, we verify that all $1,000$ obfuscated instances produce the same output as the original program for more than $1,000,000$ inputs.
To obtain a uniform distribution over varying input lengths of our cryptographic targets, we create $10,000$ random inputs for each supported input length $l \in [16;128]$. Additionally, we test a number of edge cases $\in \{0x0..0, 0xff..ff, 0x80..00, 0x00..01, 0xaa..aa, 0x55..55\}$ (or their cartesian product if two inputs are required). This amounts to a total of $1,134,068$ inputs, for which we assert equal input-output behavior.
\end{experiment}

All obfuscated binaries (both those with and without MBAs) exhibit exactly the same behavior for the $1,134,068$ inputs tested. 
\begin{experiment}[Code Coverage and Path Coverage]\label{experiment:coverage}
To further increase confidence in our correctness tests, we measure both the code coverage and the path coverage that the inputs from Experiment~\ref{experiment:correctness} achieve on the to-be-protected code both for the original program and obfuscated instances.
\end{experiment}
We find that each of the more than $1,000,000$ inputs from Experiment~\ref{experiment:correctness} achieves full code coverage and full path coverage. This ensures that our inputs cover the complete behavior of the code we obfuscate and that our obfuscation transformations have not altered this behavior.

\begin{experiment}[Overhead]\label{experiment:overhead}
For each target, we measure the average execution runtime. To this end, each target wraps the to-be-protected code in a single function, which is called $10,000$ times per input. We then execute each obfuscated binary for $1,000$ random inputs, recording the collected timings. 
We also compare the original program's disk size to the average of the obfuscated binaries.
\end{experiment}

As evident from Table~\ref{tab:eval:overhead}, the runtime overhead ranges from a factor of $301$ to $482$ compared to the original program's execution time.
While this overhead may appear excessive---also in comparison to \tigress---state-of-the-art commercial obfuscation generally imposes an even larger slowdown, up to ten times more than \loki (\cf~Table~\ref{tab:eval:overhead},~\cite{tigressblog}).
We re-run this experiment on the $1,000$ binaries without MBAs to evaluate their impact. On average, they are responsible for \textasciitilde$39\%$ of the overhead. 
Similar for the disk size, the obfuscated programs are $18$ to $51$ times larger than the original ones. Size-wise, MBAs cause \textasciitilde$33\%$ of the overhead. For further details of our MBAs' overhead, we refer to Appendix~\ref{sec:appendix:mba_overhead}.
Compared to \themida and \vmprotect, our obfuscating transformations generate almost always smaller programs, while \tigress always produces significantly smaller binaries. 
 
Overall, we conclude that our overhead is moderate in comparison to commercial state-of-the-art obfuscators. For further discussion, we refer to Section~\ref{sec:discussion}. 
\tigress' overhead is impressively small, but it falls short in providing comprehensive protection as the following experiments show.

\textbf{Case Study: \vlc with \libdvdcss.} To showcase the practical feasibility of \loki in real-world scenarios, we obfuscate the \texttt{DecryptKey} function in \libdvdcss~\cite{libdvdcss}; this component of \vlc~\cite{vlc} is responsible for decrypting the multimedia content of DVDs keys. The underlying idea is to protect the decryption algorithm from the prying eyes of crackers and protect intellectual property. 
However, the vast majority \vlc's code is irrelevant to content decryption, such that there is no need to obfuscate the whole \libdvdcss library or even the whole media player.
After obfuscating the \texttt{DecryptKey} function, which is called \emph{before} the actual media content is played, we measure the execution time of the function during initial startup, when the DVD is decrypted. We average the results over ten executions. We find that without obfuscation, the function is executed in $2,952$ nanoseconds, while with obfuscation, the decryption lasts $937,606$ nanoseconds. Overall, \loki slows down the initialization by one millisecond, a negligible cost for protecting one's intellectual property, especially if the to-be-protected function is only called in the application's startup phase.

\begin{table*}[tb]
\caption{Runtime and disk size overhead as factors relative to the non-obfuscated binaries (compiled with \texttt{O3}). \textit{(w/o = without)}}
\resizebox{\linewidth}{!}{
\begin{tabularx}{1.25\linewidth}{llRRRRRRRRRR}
\toprule
 && \multicolumn{5}{c}{Time Factor} & \multicolumn{5}{c}{Size Factor} \\
  &                                  & AES & DES & MD5 & RC4 & SHA1 & AES & DES & MD5 & RC4 & SHA1 \\
\midrule
\multirow{2}{*}{\vmprotect}
    & Virtualization & 2,489 & 1,859 & 1,982 & 1,321 & 2,524 & 37 & 21 & 40 & 44 & 40 \\
    & Ultra & 8,925 & 9,152 & 13,047 & 5,806 & 15,411 & 47 & 37 & 57 & 59 & 53 \\
\midrule
\multirow{2}{*}{\themida}
    & Tiger White & 1,388 & 622 & 203 & 240 & 552 & 58 & 38 & 58 & 58 & 59 \\
    & Dolphin Black & 11,695 & 5,052 & 2,428 & 3,634 & 8,354 & 67 & 47 & 85 & 63 & 84 \\
\midrule
\loki{}    &          & 386 & 301 & 357 & 482 & 386 & 33 &  18 &  39 &  37 &  51 \\
           & w/o MBA  & 236 & 185 & 204 & 315 & 233 & 21 &  13 &  25 &  26 &  32 \\
\midrule
\tigress{} &          & 261 &  51 & 101 &  58 & 111 &  3 &   4 &   2 &   2 &   3 \\
\bottomrule
\end{tabularx}
}
\centering%
\label{tab:eval:overhead}%
\vspace{-1.0em}
\end{table*}

\subsection{Resilience}

We evaluate whether our techniques can withstand automated deobfuscation approaches. 
To this end, we analyze the impact of syntactic and semantic attacks against the obfuscated code in the presence of both static and dynamic attackers.
We design all experiments by assuming the strongest attacker model. To this end, we test each component individually, therefore ignoring beneficial synergy effects. 
Where applicable, we first evaluate our techniques on a general design level before testing their concrete implementations. 
The former serves as universal evaluation of a technique's resilience, while the latter demonstrates that this also holds when actually implemented on the binary level.

\textbf{\lokiattack{}.} 
Fundamentally, attacking the obfuscated VM on the binary-level has two stages: (1) Identifying a specific handler within the VM, and (2) attacking (simplifying) this particular handler as far as possible. For our evaluation, especially (2) is interesting, as all our techniques focus on hardening individual handlers. As such, we develop a custom attack framework that we call \lokiattack. It is specifically tailored to the attacked obfuscators and automates the first stage: It identifies all VM handlers and provides the attacker (for each handler) with access to the handler parameters ($x$, $y$, $c$, and---for \loki---$k$). For a dynamic attacker, it also provides concrete values for these parameters. Finally, \lokiattack uses symbolic execution to obtain all code paths through the intermediate representation (IR) of the \texttt{O3}-optimized VM code that depend on an \emph{unknown} (static attacker) or \emph{known} (dynamic attacker) value of $k$ (for \loki). For each such path, an attacker can launch the actual attack on the handler (stage 2), for which \lokiattack provides a number of techniques implemented as plugins, \eg taint analysis, symbolic execution, or program synthesis. To implement \lokiattack, we use \miasm; the plugins for stage 2 are based on \triton (byte-level taint analysis), \miasm (bit-level taint analysis, backward slicing, and symbolic execution), and \syntia (program synthesis).
These plugins include costly operations (SMT solving, program synthesis, and symbolic execution), from which some may run for several days. As our evaluation consists of more than $300,000$ analysis tasks, we limit each one to $1$ hour to keep the analysis time manageable. This is a common use-case and in-line with previous work on deobfuscation~\cite{blazytko2017syntia,bardin2017backward,menguy2021xyntia}.

\subsection{Evaluation of Key Encodings}
We evaluate whether a static attacker can obtain a specific core semantics using semantic codebook checks. Note this experiment is only applicable to \loki as \tigress has no concept of key-based selection of core semantics. Assume that the function $f(x,y,c,k)$ includes $x + y$ as one of its core semantics. Then, an attacker can use an SMT solver to find a value for $k$ such that $f$ is semantically equivalent to $g(x,y,c) := x + y$.
On a technical level, we employ an approach called \emph{Counterexample-Guided Abstraction Refinement (CEGAR)}~\cite{cegar,cegarsmc} that relies on two independent SMT solvers: While SMT solver $A$ tries to find assignments for all variables (including $k$) such that $f$ and $g$ produce the same output, solver $B$ tries to find a counterexample for this value of $k$ such that $f$ and $g$ behave differently. Then, $A$ uses the counterexample as guidance. 

\begin{experiment}[Hardness of Key Encodings]\label{experiment:smt:key_encodings}
We generate $1,000$ random instances of our factorization-based key encoding and synthesize $10,000$ point functions. Then, we apply the CEGAR approach independently to both key encodings and check if the SMT solver finds a correct value for $k$.
\end{experiment}

We observe that the SMT solver found no correct key for the factorization-based encoding, but hit the 1h timeout in all cases. Considering the point function-based key encoding, \zt managed to find a value for $k$ in $6,932$ cases (\textasciitilde$69$\%). On average, it found the solution in 284s (excluding timeouts).
We conclude that an SMT solver struggles with our factorization-based key encoding, while point functions often can be solved. Recall though that point functions primarily serve to diversify and erase discernible patterns to impede pattern matching. 

\begin{experiment}[Key Encoding on Binary Level]\label{experiment:smt:low_level}
To verify if our implementation properly emits these key encodings, we generate $1,000$ binaries that contain one specific handler which includes $x + y$ as one of its core semantics. These binaries contain neither MBAs nor superoperators. 
Assuming a static attacker uses CEGAR, we check in how many cases 
the SMT solver finds a correct value for $k$.
\end{experiment}

Using \lokiattack, we obtain the handler's instructions and use our CEGAR plugin based on \zt to find a value for $k$, such that these instructions are semantically equivalent to $x + y$. While hitting the timeout in 690 cases, \zt managed to find a correct value for $k$ in 310 cases ($31$\%). The SMT solver needed, on average, 444s to find the solution (excluding timeouts).
Overall, we conclude that our key encodings indeed pose a challenge for a static attacker relying on SMT solvers. 

Note that this component is special within our system, as its approach specifically targets only static attackers. This is due to the fact that dynamic attackers can trivially observe a value for $k$. While a dynamic scenario is not always possible, another attack vector could be to offload 64-bit integer factorization to custom tools (assuming an attacker manages to locate the key encodings, which in itself is a non-trivial task given our MBAs and point functions). Thus, our key encodings can be considered to be our weakest component. However, our design assumes that an attacker can retrieve a value for $k$, but we try to make this as hard as possible. The syntactic simplification experiments show that knowledge of a key $k$ is beneficial but not sufficient to simplify any handler.

\textbf{Syntactic Simplification.}\label{sec:evaluaton:syntactic_attacks}
In the following, we evaluate whether syntactic simplification techniques---namely, forward taint analysis, backward slicing, and symbolic execution---succeed in extracting a core semantics associated with a specific key, either by trying to identify instructions not contributing to a function $f$'s output or by symbolically simplifying $f$.
We use \lokiattack as a basis and conduct the respective attack in stage 2 for both a static and a dynamic attacker.
We assume that an attacker is given a binary containing seven handlers, $f_0(x, y, c, k), \cdots, f_6(x, y, c, k))$, each containing between $3$ and $5$ core semantics. Further, each handler $f_i$ contains one predefined core semantics from the set $\{x+ y, x -y ,x \cdot y, x\land y, x\lor y, x\oplus y , x \ll y \}$ that an attacker wants to identify via syntactic simplification. As sample set for our experiments, we generate $1,000$ binaries protected by MBAs but without superoperators, amounting to $7,000$ handlers to analyze. For each binary, we use \lokiattack to extract all handlers; for each handler, \lokiattack provides us with the parameter locations (and values for the dynamic scenario) and all code paths. For each code path (a list of instructions), we then use the respective stage 2 plugin.
We apply the following experiments also to $7,000$ \tigress handlers (with disabled superoperators), respectively.

\begin{table}[tp]
    \caption{Statistics for backward slicing and forward taint analysis (TA), averaged over $7,000$ handlers. Unmarked instruction can be removed as irrelevant. \ \ \textit{(Unmark. = not tainted / not sliced; \ Dyn. = dynamic attacker)}}
    \centering
\resizebox{\linewidth}{!}{
\begin{tabular}{llrrrrrr}
\toprule
&& \multicolumn{2}{c}{Byte-level TA} & \multicolumn{2}{c}{Bit-level TA} & \multicolumn{2}{c}{Slicing} \\
&&             Static & Dyn. & Static & Dyn. & Static & Dyn.\\ 
\midrule
\multirow{3}{*}{\rotatebox[origin=c]{90}{\loki}} &IR paths            & 1,950    &     199 &  1,451    &   168    & 1,656    &   179    \\
                         &Unmark.       & 17.49\%  & 17.50\% &   17.61\% &  17.62\% &  5.49\%  &   7.57\% \\
                         &Time [s]            &   556    &     58  &    710    &    78    &   630    &    67 \\
\addlinespace[0.2em]
\midrule
\addlinespace[0.6em]
\multirow{3}{*}{\rotatebox[origin=c]{90}{\parbox[c]{3.5em}{\centering \tigress}}}&IR paths            & \multicolumn{2}{c}{1}  &        \multicolumn{2}{c}{1}    &  \multicolumn{2}{c}{1}   \\
                         &Unmark.       & \multicolumn{2}{c}{44.70\%} &   \multicolumn{2}{c}{44.70\%} &  \multicolumn{2}{c}{22.35\%} \\
                         &Time [s]            &   \multicolumn{2}{c}{1.3}   &   \multicolumn{2}{c}{1.6}   &   \multicolumn{2}{c}{1.4}  \\
\addlinespace[0.3em]
\bottomrule
\end{tabular}
}%
\label{tab:eval:taint_and_slicing}%
\vspace{-1.5em}
\end{table}

\begin{experiment}[Forward Taint Analysis]\label{experiment:taint_analysis}
For each of the $7,000$ handlers, we conduct a forward taint analysis with byte-level and bit-level granularity. The former is based on \triton, while the more precise bit-level taint analysis is implemented on top of \miasm. In general, higher precision is expected to produce fewer false positives and result in fewer tainted instructions. Recall that an attacker's goal is to identify all instructions that do not belong to the core semantics associated with a specific key. Using taint analysis, an attacker can remove all instructions not depending on $x$, $y$, $c$, or $k$ (in a dynamic scenario: on a \emph{concrete} value for $k$).%
\end{experiment}
The resulting data is shown in Table~\ref{tab:eval:taint_and_slicing} (where \emph{unmarked} instructions refer to instructions that are \emph{not} tainted, \ie instructions that can be removed).
The results show two interesting insights: First, the granularity has negligible impact on the results (difference of $0.12\%$).
Second, the number of tainted instructions is almost equal for a static and a dynamic attacker. This is surprising as for \loki the number of visited paths in the IR's control-flow graph is significantly lower in the dynamic setting. Intuitively, this means a dynamic attacker has better chances of removing more instructions. However, our results show that the sole benefit of a dynamic attacker is spending less time per handler.
In numbers, an attacker is always able to only remove about \textasciitilde$18\%$ of a handler's assembly instructions. Manually inspecting the instructions not tainted revealed that these can always be put into two categories: Either they are part of the inlined VM dispatcher that is responsible for loading the next handler (which is independent of the current handler's semantics), or it is an instruction loading a constant value before it interacts with tainted instructions (comparable to Example~\ref{example:taint}).
To summarize, forward taint analysis fails to remove a single instruction that is related to the core semantics or key encodings.
For \tigress, on the contrary, only one IR path exists, meaning the handlers are short and simplistic in nature. No difference between bit and byte-wise taint analysis exists; overall,  an attacker succeeds in removing 45\% of instructions---significantly more than for \loki.

\begin{experiment}[Backward Slicing]
Besides forward taint analysis, an attacker can use backward slicing to identify all instructions that contribute to a handler's output. We again consider both a static and dynamic attacker trying to reduce each handler to the core semantics associated with a specific $k$ by removing as many unrelated instructions as possible. Our backward slicing approach is based on \miasm and operates on the same $7,000$ handlers as \ex{experiment:taint_analysis}. %
\end{experiment}
The results are denoted in Table~\ref{tab:eval:taint_and_slicing}, where an \emph{unmarked} instruction refers to an instruction that was not sliced, \ie it does not contribute to the output.
Other than for taint analysis, a dynamic attacker has a slight advantage compared to a static attacker ($2.08\%$), as they slice slightly fewer instructions. 
While the static attacker marks all instructions but the inlined dispatcher, our manual inspection revealed that dynamic analysis skips some IR paths depending on irrelevant key values.
Compared to forward taint analysis, backward slicing marks more instructions, \ie it removes fewer instructions (\textasciitilde$7\%$ vs.\ \textasciitilde$18\%$). This is due to the backward-directed nature of the approach, which allows it to slice instructions loading constant values. We conclude that backward slicing is technically more precise than forward taint analysis, but fails to remove instructions belonging to the core semantics or key encodings. For \tigress, slicing succeeds to remove significantly more instructions, however, less than taint analysis. This is again due to the loading of constant values.

\begin{table}[tp]
      \centering
        \caption{Symbolic execution for semantic depth $3$ and $5$, each averaged over $7,000$ handlers. \ \ \textit{(Simplified = percentage of handlers simplified)}}
\resizebox{\linewidth}{!}{
\begin{tabular}{llrr@{\hskip 2em}rr}
\toprule
&& \multicolumn{2}{c}{Depth 3} & \multicolumn{2}{c}{Depth 5} \\
&&  Static & Dynamic                      &    Static & Dynamic                        \\ 
\midrule
\multirow{3}{*}{\loki{}}    &IR paths          & 4,960 &     559 &  5,450 &     703 \\
                            &Simplified              &   0\% & 17.93\% &    0\% & 14.64\% \\
                            &Time [s]          &    -- &    94 &     --  &    168 \\
\midrule
\multirow{3}{*}{\tigress}   &IR paths          &    \multicolumn{2}{c}{1} &  \multicolumn{2}{c}{--} \\
                            &Simplified      &   \multicolumn{2}{c}{30.61\%} &    \multicolumn{2}{c}{--} \\
                            &Time [s]          &   \multicolumn{2}{c}{1.4}&     \multicolumn{2}{c}{--} \\
\bottomrule
\end{tabular}
}
\centering
\label{tab:eval:symbolic_execution}
\vspace{-2em}
\end{table}

\begin{experiment}[Symbolic Execution]\label{experiment:symbolic_execution}
Besides removing instructions not contributing to the output, an attacker can use symbolic execution to extract a handler's core semantics. To this end, a symbolic executor uses simplification rules to syntactically simplify the handler's semantics as much as possible.
We use the same $7,000$ handlers as \ex{experiment:taint_analysis}. We analyze each handler independently with \miasm's symbolic execution engine and measure whether it can be simplified to the original core semantics. %
\end{experiment}

We model both a static and more powerful dynamic attacker. In the latter scenario, the attacker observes a value for $k$ and thus can nullify all core semantics not related to this specific $k$. Hence, they obtain a much simpler expression containing only the desired core semantics, albeit in syntactically complex form (due to MBAs).
Recall that for the $7,000$ handlers, the semantic depth of the core semantics is always $3$ (\eg $x + y$). This intentionally weakens resilience, as superoperators with a higher depth naturally increase both semantic and syntactic complexity. To show this, we create another $7,000$ handlers ($1,000$ binaries à $7$ core semantics) with a semantic depth of $5$ (\eg $x + y + c$) and repeat this experiment. We cannot create handlers of depth 5 for \tigress, as it is not possible to explicitly set the handler's semantic depth.

All results are shown in Table~\ref{tab:eval:symbolic_execution}. Notably, a static attacker fails to simplify any of \loki's handlers. Without knowing a value for $k$, an attacker has to analyze the expression containing \emph{all} key encodings and their associated core semantics. In other words, an attacker fails to nullify irrelevant core semantics. To significantly simplify the handler, a static attacker has to find a valid key first (reducing the problem to \ex{experiment:smt:low_level}).
A dynamic attacker, on the other hand, only has to simplify the MBAs as they already identified the core semantics associated with the key. For depth $3$, they succeed in removing all MBAs for \textasciitilde$18\%$ of \loki's handlers, while, for \tigress, \textasciitilde$31\%$ of the handlers can be simplified. In other words, an attacker can simplify significantly more handler for \tigress than for \loki.
For a more realistic depth of $5$---subsequent experiments show \textasciitilde$80\%$ of \loki's handlers are at least of depth $5$---the attacker's success rate is even lower, namely \textasciitilde$15\%$. 
This percentage implies that a number of expressions can be simplified regardless of the higher base depth. This may be the case, \eg when the random combination of applied MBAs cancels itself out.
Still, this demonstrates our synergy effects are indeed helpful to prevent an attacker from symbolically simplifying the core semantics, leaving them with a complex MBA that conceals the actual semantics. 

We conclude that our MBAs are successful in thwarting symbolic execution, one of the most powerful deobfuscation attacks. For a more detailed analysis of how a user of \loki can trade performance against reducing the attacker's success chances even further (to $6.79\%$), refer to Appendix~\ref{sec:appendix:se_mba_bounds}.

\begin{experiment}[Diversity of MBAs]\label{experiment:mba_diversity}
An attacker tasked with removing such MBAs may investigate whether a diverse number of expressions exists for the same core semantics. If this is not the case, they can manually analyze each MBA and extend the symbolic executor's limited set of simplification rules by rewriting rules to \enquote{undo} specific MBAs. 
To this end, we assume a dynamic attacker that already symbolically simplified the expression as far as possible without any MBA-specific simplification rules. We do this for each handler type (recall that the $7,000$ handlers of depth 3 consist of $7$ different core semantics à $1,000$ handlers) and then analyze how many different, unique MBA expressions exist.
\end{experiment}

Our analysis reveals that, in summary, \loki generates $5,482$ unique MBAs for the $7,000$ expressions analyzed ($78.31\%$), while \tigress creates only $16$  (\textasciitilde$0.23\%$) unique MBAs. Thus, an attacker adding 16 rules to their symbolic executor could simplify all core semantics.
This difference can be explained by the fact that \tigress uses only a few handwritten rules to create MBAs, while \loki features a generic approach to synthesize MBAs. %
To further highlight the difference between both approaches, we repeat this experiment for another set of $7,000$ handlers---created in the same configuration but with different random seeds---and calculate the intersection of unique MBAs. \tigress re-uses exactly the same $16$ MBAs, while \loki re-uses $109$ expressions but generates $5,299$ new unique MBAs (\ie $10,781$ unique MBAs in total). 
Creating simplification rules specific to \loki is a tedious task (given the high number of unique MBAs) that does not pay off when analyzing other obfuscated instances. 
For a discussion of what an attacker can achieve when they are in possession of \emph{all} available MBA rewriting rules, refer to Section~\ref{sec:discussion}. %
We conclude that \loki's MBAs are superior to state-of-the-art approaches relying on a small number of hardcoded MBAs, both in terms of resilience and diversity.

\textbf{State-of-the-art MBA Deobfuscation.} A number of approaches for MBA simplification have been proposed, most notably \sspam~\cite{eyrolles2016defeating}, \arybo~\cite{guinet2016arybo}, \neureduce~\cite{feng2020neureduce}, and \mbablast~\cite{liu2021mbablast}. The deployed techniques range from pattern matching-based simplification over machine learning to algebraic simplification. Regardless of the underlying technique, they all share one major drawback: They expect the MBAs to be available on the source code level in form of a formula, such as \enquote{x + y - y}, rather than dealing with them on the binary level. As a consequence, these deobfuscation tools lack support for MBAs using different bit sizes and operations such as zero-extension or sign-extension. Furthermore, they assume that the MBAs are free of constants and more complex arithmetic operations, such as multiplication or left-shifts. In contrast to these limitations and assumptions, \loki's MBAs not only employ all these operators but also contain constants, thus making a fair, direct evaluation of our MBAs contained in binaries difficult. 
To avoid these pitfalls, we use \loki's term rewriter to generate simpler MBAs---called \textit{artificial} MBAs---and emit them as a formula rather than deploying them in the binary. We make the following artificial restrictions: \begin{inparaenum}[(a)] We \item emit no constants, \item do not intertwine the MBAs with the key encoding, \item remove any information (or operation) relating to size casts, and \item avoid complex, unsupported operations such as multiplications or left-shifts\end{inparaenum}. Instead, the resulting MBAs are a formula containing only the following operations $\{+, -, \land, \lor, \oplus\}$. While this significantly weakens \loki's MBAs, aforementioned state-of-the-art MBA deobfuscation techniques can now process these artificial MBAs, allowing a fair evaluation.
For \neureduce, we use the Gated Recurrent Unit (GRU)-based Long Short-Term Memory (LSTM) model provided by the authors. We further adapt \mbablast to recursively attempt simplification for subexpressions. \mbablast cannot deal with nested arithmetic expressions; only expressions on the root level of the expression's abstract syntax tree may contain arithmetic operators. All subexpressions must consist purely of Boolean operators. As our MBAs are highly nested, we apply the respective tool recursively on each subexpression until it cannot simplify the expression any longer. We considered evaluating \arybo~\cite{guinet2016arybo}; however, we noticed it does not terminate for 64-bit expressions within one hour. Further, \arybo outputs truth tables in form of expressions representing the relations between different bit positions. Its goal is aiding a human analyst rather than automated simplification. Thus, we exclude it from the following experiment.
As a baseline, we port our deobfuscation tooling, \lokiattack with the SE plugin, to the source level: We first use aggressive compiler optimizations (\enquote{\texttt{-O3}}) to simplify the MBA and then---as a stage 2 plugin---symbolically summarize it using \miasm's symbolic execution engine. This is the same approach as has been used for the previous experiments.
\begin{experiment}[MBA Formula Deobfuscation]\label{experiment:mba_text_level}
For each core semantics from the set $\{x+ y, x -y , x\land y, x\lor y, x\oplus y \}$, we use \loki to generate $1,000$ artificially simplified MBAs on the source code level. We do this for each recursive term rewriting bound from $[1,30]$ (during normal operation, \loki's rewriting bound is randomly chosen between $[20,30]$). In summary, we generate $5,000$ MBAs per rewriting bound, \ie $150,000$ obfuscated expressions in total. We then pass each MBA to the deobfuscation tools \mbablast, \neureduce, \sspam, and \lokiattack and observe how many expressions they can simplify to the ground truth. 
\end{experiment}
\begin{figure}[tp]
\centering
    \includegraphics[width=.99\linewidth]{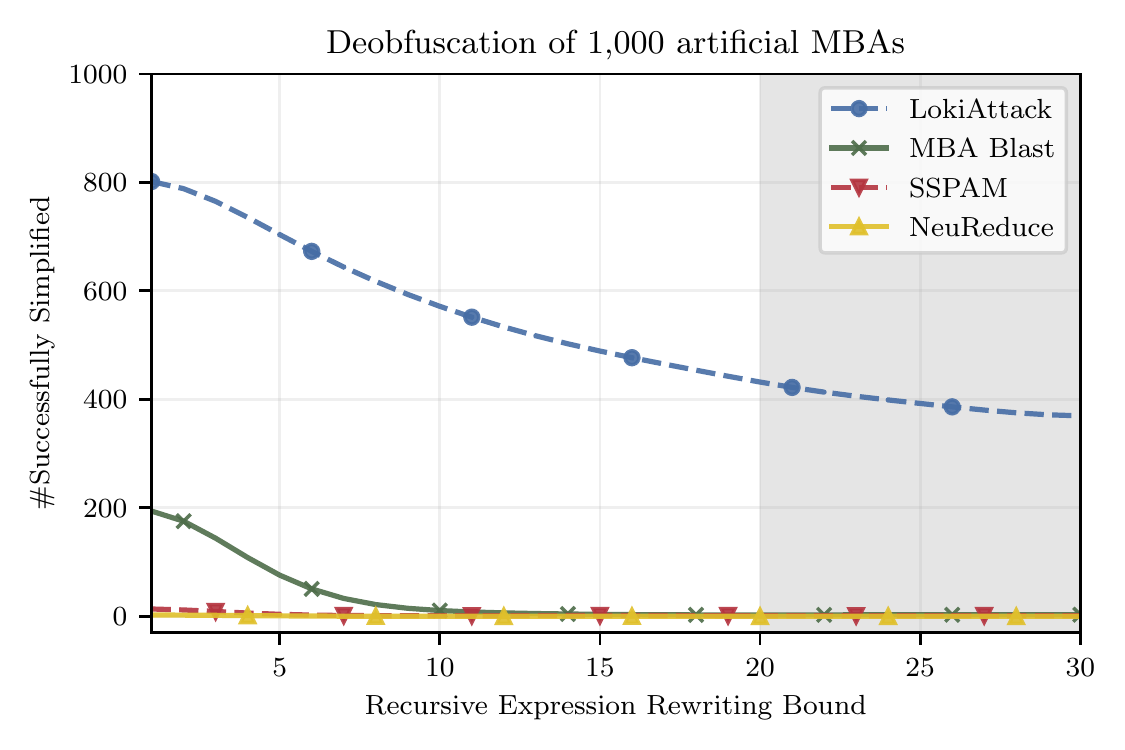}
    \caption{Number of artificial MBAs that have been successfully simplified (averaged over 5 different core semantics). The gray-shaded area marks the recursive rewriting bounds randomly picked by \loki for our regular MBAs.}
    \label{fig:mba_text_level_deobfuscation}
    \vspace{-0.6em}
\end{figure}
The number of simplified expressions, averaged over the five different core semantics, are depicted in Figure~\ref{fig:mba_text_level_deobfuscation}. 
As the data shows, our custom deobfuscation tooling, \lokiattack, significantly outperforms all state-of-the-art deobfuscation techniques. \neureduce can only simplify a handful of expressions in total. One limitation is that it can only work with inputs up to $100$ characters; however, our artificial MBAs with a rewriting bound of 20 contain, on average, $7,960$ characters (after removing all whitespaces). We have tried a similar approach as we have employed for \mbablast, however, found it does not improve its accuracy. Studying their dataset used to train the model, we believe that their approach heavily overfits on the training data, a set of simple and short (on average 75 characters without whitespace) MBAs. \sspam fails to deal with the highly recursive nature of our MBAs, frequently hitting the stack recursion limit. \mbablast performs better and manages to simplify a number of simple MBAs. However, the success rate of all tools decreases with a higher term rewriting bound. \loki's default is to use a random recursive rewriting bound between $20$ and $30$, for which all but \lokiattack fail to simplify basically any MBA. For example, \mbablast simplifies only $157$ of $55,000$ MBAs for \loki's recursive rewriting bounds, $[20,30]$. While the success rate of \lokiattack may seem high, recall that we artificially weakened these MBAs by excluding a number of operations and removing all constants; Experiment~\ref{experiment:symbolic_execution} evaluates how \lokiattack with the symbolic execution plugin performs on our regular MBAs.

\textbf{Semantic Attacks.}\label{sec:evaluation:semantic_attacks}
Semantic attacks such as program synthesis exploit the low semantic depth of individual core semantics. We evaluate the impact of our superoperators on these attacks. First, we analyze the average semantic complexity of core semantics with and without superoperators. Then, we perform a high-level experiment to measure the general limits of synthesis-based approaches. Finally, we demonstrate that our superoperators withstand synthesis-based attacks on the binary level. Note that we only consider a dynamic attacker in the following, as knowing a value for $k$ is a prerequisite for any reasonable semantic attack. A static attacker would only learn random behavior, as the key encodings are only valid for a predefined set of keys.

\begin{experiment}[Complexity of Core Semantics]\label{experiment:core_semantics_complexity}
To evaluate our superoperators' distribution and their impact on the complexity of core semantics, \ie their semantic depth,
we create $1,000$ obfuscated binaries \emph{without} superoperators as a baseline for each benchmarking target (\cf Section~\ref{sec:benchmarking}) and $1,000$ binaries \emph{with} superoperators. We compare the two sets on the average number of unique core semantics and their semantic depths. To simplify evaluation, no MBAs are used.
\end{experiment}

\begin{figure}[tp]
\centering
    \includegraphics[width=.9\linewidth]{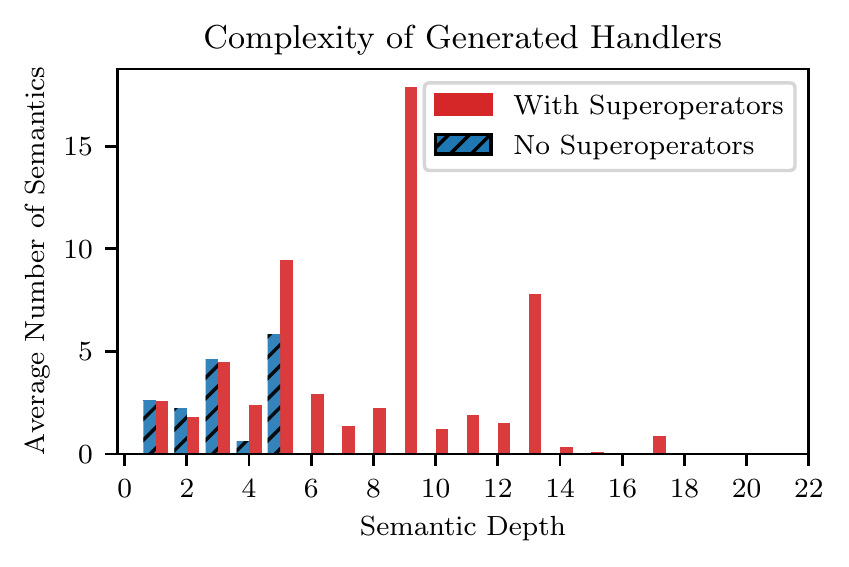}
    \caption{The distribution of core semantics with and without superoperators.}
    \label{fig:superoperator_depths}
    \vspace{-0.6em}
\end{figure}
Without superoperators, each binary on average contains $15.8$ core semantics. With superoperators, this number increases to $58.8$. Additionally, Figure~\ref{fig:superoperator_depths} shows that superoperators have a significantly higher semantic depth, usually ranging from $5$ to $13$ with a clearly visible peak at depth $9$. 
Compared to obfuscation without superoperators, where only a few core semantics with semantically low complexity are used, superoperators increase the number of unique core semantics and their semantic depth notably.
This makes the task of synthesizing semantics more difficult.

\begin{experiment}[Limits of Program Synthesis]\label{experiment:synthesis_limits}
We evaluate how the success rates of program synthesis relate to semantic complexity. We generate $10,000$ random expressions for each semantic depth between $1$ and $20$ and measure how many of them can be synthesized successfully.
Modeling our function $f$, we use \syntia's grammar~\cite{blazytko2017syntia} to generate random expressions depending on three variables. 
Based on the authors' guidance, we set \syntia's configuration vector to $(1.5,~50000,~20,~0)$ and use it to synthesize each expression. %
\end{experiment}
\begin{figure}[tp]
\centering
    \includegraphics[width=.9\linewidth]{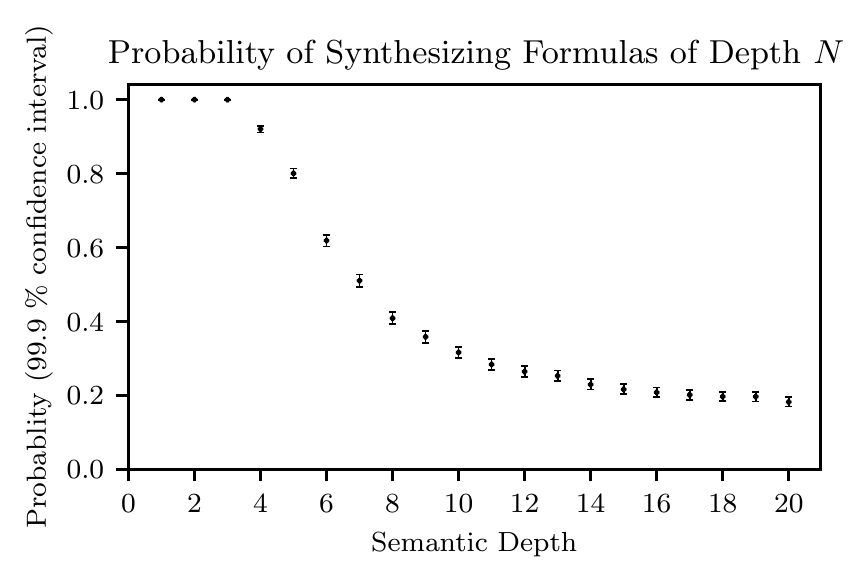}
    \caption{The probability to synthesize a valid candidate for formulas of depth $N$. The error bars are calculated as the 99.9\% confidence interval for the true probability.}
    \label{fig:synthesis_limits}
    \vspace{-2em}
\end{figure}

Figure~\ref{fig:synthesis_limits} shows that simple expressions can be synthesized quite easily; at a semantic depth of $7$, only \textasciitilde$50\%$ can be synthesized. For larger semantic depths, it becomes increasingly unlikely to synthesize expressions. Given our results from \ex{experiment:core_semantics_complexity}, we conclude that our superoperators produce core semantics of sufficient depth to impede program synthesis.

\begin{experiment}[Superoperators on Binary Level]\label{experiment:synthesis}
To evaluate the impact of program synthesis, we assume a dynamic attacker has extracted a handler's core semantics (for \loki: associated with a known value of $k$). They then use \syntia---configured as in \ex{experiment:synthesis_limits}---to learn an expression having the same input-output behavior.
We create $400$ obfuscated binaries (without MBAs, with superoperators) for each benchmarking target (\cf Section~\ref{sec:benchmarking}), randomly pick $10,000$ core semantics and measure \syntia's success rate.
\end{experiment}
Overall, using \syntia as a stage 2 plugin on top of \lokiattack, we managed to synthesize $1,888$ (\textasciitilde$19\%$) of \loki's expressions and $6,779$ (\textasciitilde$68\%$) of \tigress' expressions. On average, it took $157$s to synthesize an expression for \loki and $144$s for \tigress. The results show that---while both designs employ superoperators---it is crucial how these superoperators are crafted.
As outlined in Section~\ref{paragraph:superoperators}, \tigress usually includes independent core semantics, allowing the attacker to split the superoperators into multiple smaller synthesis tasks, each of low semantic complexity. On the other hand, \loki's design ensures that its superoperators cannot be split into smaller tasks but have high semantic depths.
In summary, \loki is the first obfuscation design showing sufficient protection against program synthesis, an attack vector all state-of-the-art obfuscators fail to account for.
Given \loki's synergy effects and high resilience against syntactic simplification approaches, semantic deobfuscation techniques remain an attacker's last resort. However, even when using program synthesis, arguably the strongest semantic attack, an attacker can only recover less than a fifth of \loki's core semantics.

\section{Discussion}\label{sec:discussion}

\textbf{Overhead.} Table~\ref{tab:eval:overhead} indicates that the overhead of code obfuscation is generally excessive. However, this cost is accepted in practice because only small, critical parts of the whole program need to be protected (\eg proprietary algorithms, API accesses, or licensing-related code). As a result, the overhead has to be seen in relation to the whole program. 
As our case study shows for \libdvdcss, using obfuscation only for critical, well-chosen code parts has no negative impact on the usability of the respective program (here \vlc).

\textbf{MBA Database.}
Assuming an attacker intends to symbolically simplify MBAs, they may benefit from using a lookup table mapping complex MBAs to simpler expressions. This approach is effective for state-of-the-art obfuscators such as \tigress that only use a limited number of hardcoded rewriting rules (\cf \ex{experiment:mba_diversity}).
\loki, in contrast, is the first obfuscator that employs a generic approach to synthesize highly diverse MBAs, resulting in a large number of MBAs (stored in a database for performance reasons). Users of \loki can keep their MBA database (including the synthesis limit up to which MBAs were synthesized) private. In fact, users could choose arbitrary lower and upper limits as well as completely different grammars to create an MBA database. Without knowing the parameters, a re-creation of the database is not feasible. 
That said, even in cases where an attacker is in possession of the MBA database, there is no straightforward process to reverse the recursively generated expression (\cf Section~\ref{sec:design:mba_synthesis}).

\textbf{Attacker Model.} Our evaluation assumes a strong attacker model with significant domain knowledge and access to all kinds of static and dynamic analyses. In practice, an attacker is often weaker. Especially without prior knowledge about the given obfuscation techniques, the usage of additional techniques (\eg VM bytecode blinding~\cite{blazytko2017syntia} or range dividers~\cite{banescu2016code}) and other countermeasures (\eg self-modifying code~\cite{obf-selfmodify} or anti-debugging techniques~\cite{gagnon2007software}) complicates analysis.

\textbf{Human Attacker.} Ultimately, code obfuscation schemes are usually broken by human analysts~\cite{rolles2009unpacking}. 
This is partly because humans excel at recognizing patterns and adapt to the given obfuscation~\cite{dang2014practical}. Collberg~\etal~\cite{collberg1997taxonomy} define \emph{potency} to denote how \emph{confusing} an obfuscation is for a human analyst. Due to the difficulty of measuring a human's capability with regard to deobfuscation, we restrict our evaluation to automated attacks. We argue that without automated techniques, analysis becomes subjectively harder. Nevertheless, we believe that pattern matching might be the most potent attack on our approach. While we use a fixed structure, %
we argue that our MBAs remove identifiable patterns. Still, we are not aware of an adequate way of measuring this. However, even if we assume that a human attacker breaks one obfuscated instance, other instances remain hard. This is as our design samples from large search spaces for its critical components, providing significant diversity for MBAs and superoperators. In summary, we expect \loki to perform reasonably well against human attackers even if this cannot be easily quantified.

\section{Related Work}
Over the years, a large number of obfuscation techniques were proposed~\cite{collberg1997taxonomy,zhou2007information,obf-general,obf-probfuscation,obf-vmbased,obf-controlflow,obf-selfmodify,banescu2016code,wang2011linear,ollvm,ollivier2019howto,nakanishi2020intertwining,borrello2021hiding}. Many of these techniques are orthogonal to our work and focus on one specific transformation. For an overview over the field of obfuscation, we refer the interested reader to the overview by Banescu and Pretschner~\cite{banescu2018tutorial}. 
In the following, we discuss techniques closest to our work.

\textbf{MBA.}
Zhou~\etal introduced the concept of Mixed Boolean-Arithmetic (MBA) to hide constants and calculations within complex expressions. While conceptually simple, this approach proved effective against many analysis techniques, such as symbolic execution. As a consequence, a number of approaches towards deobfuscating MBAs were proposed, including
pattern matching (\sspam~\cite{eyrolles2016defeating}), 
symbolic simplification using a Boolean expression solver (\arybo~\cite{guinet2016arybo}),
program synthesis (\syntia~\cite{blazytko2017syntia}, \xyntia~\cite{menguy2021xyntia}, \qsynth~\cite{david2020qsynth}),
machine learning (\neureduce~\cite{feng2020neureduce}), and
algebraic simplification (\mbablast~\cite{liu2021mbablast}).
While those techniques are effective against common MBAs, \loki's generic approach to synthesize diverse MBAs produces expressions resilient against such attacks (\cf~Section~\ref{sec:evaluation}).

\textbf{VM Obfuscation.}
Our prototype implementation \loki uses a VM-based architecture to showcase our techniques. However, we make no attempt at obfuscating the VM structure itself, which we consider orthogonal to our work. 
Examples for such work include \emph{virtual code folding}, where the mapping between opcodes and individual handlers is obfuscated to impede static analyses~\cite{suk2020vcf,lee2018vodka,xue2018exploiting,cheng2019dynopvm}. While they use dynamic keys to determine the next handler, we use keys within our handlers to select a specific core semantics. 
With regard to deobfuscation, approaches such as \vmhunt~\cite{xu2018vmhunt}, \vmattack~\cite{kalysch2017vmattack}, and others~\cite{kinder2012towards,sharif2009automatic} may succeed in reconstructing \loki's VM structure (similar to \lokiattack). However, they cannot recover individual handler semantics, since they rely on techniques such as symbolic execution and backward slicing, for which \loki is resilient against by design.

\textbf{Thwarting Symbolic Execution.}
With regard to thwarting symbolic execution-based deobfuscation approaches, early work by Sharif~\etal~\cite{sharif2008impeding} already proposed key-based encodings to make path exploration infeasible. Later approaches extend on this work by introducing multi-level opaque predicates (so-called \emph{range dividers})~\cite{banescu2016code} or artificial loops~\cite{ollivier2019howto}.
\loki extends these ideas: it does not only make path exploration infeasible, but also prevents symbolic simplification attacks due to its MBAs.

\textbf{Thwarting Program Synthesis.}
Program synthesis is one of the most powerful attack vectors~\cite{blazytko2017syntia,menguy2021xyntia}. 
Concurrent work~\cite{menguy2021xyntia} proposes a search-based program synthesis approach outperforming \syntia. However, the authors note that merging handlers and increasing a handler's semantic complexity proved effective in thwarting such attacks. This is in line with our evaluation.

\section{Conclusion}

In this paper, we present and extensively evaluate a set of novel and generic obfuscation techniques that, in combination, succeed to thwart automated deobfuscation attacks. Our four core techniques include a novel and generic approach to synthesize and formally verify MBAs of arbitrary complexity, overcoming the limits imposed by using hardcoded rules. We further include a new approach to increase obfuscation's semantic complexity, based on an investigation of the limits of program synthesis.
In conclusion, we show that a comprehensive and effective intellectual property protection can be achieved without excessive overheads.

\urlstyle{tt}

\bibliographystyle{plain}
\bibliography{bib}

\begin{thebibliography}{10}

\bibitem{banescu2016code}
Sebastian Banescu, Christian Collberg, Vijay Ganesh, Zack Newsham, and
  Alexander Pretschner.
\newblock {Code Obfuscation against Symbolic Execution Attacks}.
\newblock In {\em Annual Computer Security Applications Conference (ACSAC)},
  2016.

\bibitem{banescu2017predicting}
Sebastian Banescu, Christian Collberg, and Alexander Pretschner.
\newblock {Predicting the Resilience of Obfuscated Code Against Symbolic
  Execution Attacks via Machine Learning}.
\newblock In {\em USENIX Security Symposium}, 2017.

\bibitem{banescu2018tutorial}
Sebastian Banescu and Alexander Pretschner.
\newblock {A Tutorial on Software Obfuscation}.
\newblock {\em Advances in Computers}, 108:283--353, 2018.

\bibitem{bansal2006automatic}
Sorav Bansal and Alex Aiken.
\newblock {Automatic Generation of Peephole Superoptimizers}.
\newblock In {\em ACM Sigplan Notices}, 2006.

\bibitem{bardin2017backward}
S{\'e}bastien Bardin, Robin David, and Jean-Yves Marion.
\newblock {Backward-Bounded {DSE}: Targeting Infeasibility Questions on
  Obfuscated Codes}.
\newblock In {\em IEEE Symposium on Security and Privacy}, 2017.

\bibitem{barhelemy2016binary}
Lucas Barhelemy, Ninon Eyrolles, Guena{\"e}l Renault, and Rapha{\"e}l Roblin.
\newblock {Binary Permutation Polynomial Inversion and Application to
  Obfuscation Techniques}.
\newblock In {\em ACM Workshop on Software PROtection (SPRO)}, 2016.

\bibitem{blazytko2017syntia}
Tim Blazytko, Moritz Contag, Cornelius Aschermann, and Thorsten Holz.
\newblock {Syntia: Synthesizing the Semantics of Obfuscated Code}.
\newblock In {\em USENIX Security Symposium}, 2017.

\bibitem{borrello2021hiding}
Pietro Borrello, Emilio Coppa, and Daniele~Cono D'Elia.
\newblock {Hiding in the Particles: When Return-Oriented Programming Meets
  Program Obfuscation}.
\newblock In {\em Conference on Dependable Systems and Networks (DSN)}, 2021.

\bibitem{sha1implementation}
Jasin Bushnaief.
\newblock {SHA-1}.
\newblock \url{https://gist.github.com/rverton/a44fc8ca67ab9ec32089}, 2016.

\bibitem{cavallaro2007anti}
Lorenzo Cavallaro, Prateek Saxena, and R~Sekar.
\newblock {Anti-Taint-Analysis: Practical Evasion Techniques against
  Information Flow-based Malware Defense}.
\newblock Technical report, Secure Systems Lab, Stony Brook University, 2007.

\bibitem{miasm-github}
{CEA IT Security}.
\newblock {Miasm -- Reverse Engineering Framework}.
\newblock \url{https://github.com/cea-sec/miasm}.

\bibitem{antidebugging2}
Ping Chen, Christophe Huygens, Lieven Desmet, and Wouter Joosen.
\newblock {Advanced or Not? {A} Comparative Study of the Use of Anti-Debugging
  and Anti-{VM} Techniques in Generic and Targeted Malware}.
\newblock In {\em IFIP International Conference on ICT Systems Security and
  Privacy Protection}, pages 323--336, 2016.

\bibitem{antidebugging}
Xu~Chen, Jon Andersen, Z.~Morley Mao, Michael Bailey, and Jose Nazario.
\newblock {Towards an Understanding of Anti-Virtualization and Anti-Debugging
  Behavior in Modern Malware}.
\newblock In {\em Conference on Dependable Systems and Networks (DSN)}, pages
  177--186. IEEE, 2008.

\bibitem{cheng2019dynopvm}
Xiaoyang Cheng, Yan Lin, Debin Gao, and Chunfu Jia.
\newblock {DynOpVm: VM-based Software Obfuscation with Dynamic Opcode Mapping}.
\newblock In {\em International Conference on Applied Cryptography and Network
  Security}, 2019.

\bibitem{tigressblog}
{Christian Collberg}.
\newblock {Performance vs. Security}.
\newblock \url{https://tigress.wtf/blog.html}.

\bibitem{cegar}
Edmund Clarke, Orna Grumberg, Somesh Jha, Yuan Lu, and Helmut Veith.
\newblock {Counterexample-guided Abstraction Refinement}.
\newblock In {\em International Conference on Computer Aided Verification},
  2000.

\bibitem{cegarsmc}
Edmund Clarke, Orna Grumberg, Somesh Jha, Yuan Lu, and Helmut Veith.
\newblock {Counterexample-guided Abstraction Refinement for Symbolic Model
  Checking}.
\newblock {\em Journal of the ACM (JACM)}, 50(5):752--794, 2003.

\bibitem{tigress}
Christian Collberg.
\newblock {The {Tigress C} Diversifier/Obfuscator}.
\newblock \url{http://tigress.cs.arizona.edu/}.

\bibitem{collberg1997taxonomy}
Christian Collberg, Clark Thomborson, and Douglas Low.
\newblock {A Taxonomy of Obfuscating Transformations}.
\newblock Technical report, Department of Computer Science, The University of
  Auckland, New Zealand, 1997.

\bibitem{collberg1998manufacturing}
Christian Collberg, Clark Thomborson, and Douglas Low.
\newblock {Manufacturing Cheap, Resilient, and Stealthy Opaque Constructs}.
\newblock In {\em ACM Symposium on Principles of Programming Languages (POPL)},
  1998.

\bibitem{coogan2011deobfuscation}
Kevin Coogan, Gen Lu, and Saumya Debray.
\newblock {Deobfuscation of Virtualization-obfuscated Software: A
  Semantics-Based Approach}.
\newblock In {\em ACM Conference on Computer and Communications Security
  (CCS)}, 2011.

\bibitem{cytron1989ssa}
Ron Cytron, Jeanne Ferrante, Barry~K. Rosen, Mark~N. Wegman, and F.~Kenneth
  Zadeck.
\newblock {An Efficient Method of Computing Static Single Assignment Form}.
\newblock In {\em ACM Symposium on Principles of Programming Languages (POPL)},
  1989.

\bibitem{dang2014practical}
B.~Dang, A.~Gazet, E.~Bachaalany, and S.~Josse.
\newblock {\em {Practical Reverse Engineering: x86, x64, ARM, Windows Kernel,
  Reversing Tools, and Obfuscation}}.
\newblock Wiley, 2014.

\bibitem{danicic2018static}
Sebastian Danicic and Michael~R. Laurence.
\newblock {Static Backward Slicing of Non-Deterministic Programs and Systems}.
\newblock {\em {{ACM} Transactions on Programming Languages and Systems
  (TOPLAS)}}, 40(3):11:1--11:46, 2018.

\bibitem{david2020qsynth}
Robin David, Luigi Coniglioi, and Mariano Ceccato.
\newblock {QSynth -- A Program Synthesis based Approach for Binary Code
  Deobfuscation}.
\newblock In {\em Symposium on Network and Distributed System Security (NDSS),
  Workshop on Binary Analysis Research}, 2020.

\bibitem{denuvo}
{Denuvo Software Solutions GmbH}.
\newblock {{Denuvo} Anti-Tamper}.
\newblock \url{http://www.denuvo.com}.

\bibitem{peach}
Michael Eddington.
\newblock {Peach Fuzzer: Discover Unknown Vulnerabilities}.
\newblock \url{https://www.peach.tech/}.

\bibitem{snapchat_obfuscation}
Abdelrahman Eid.
\newblock {Reverse Engineering Snapchat (Part I): Obfuscation Techniques}.
\newblock \url{https://hot3eed.github.io/snap_part1_obfuscations.html}.

\bibitem{eyrolles2017dissertation}
Ninon Eyrolles.
\newblock {\em {Obfuscation with Mixed Boolean-Arithmetic Expressions:
  Reconstruction, Analysis and Simplification Tools}}.
\newblock PhD thesis, {Université de Versailles Saint-Quentin-en-Yvelines},
  2017.

\bibitem{eyrolles2016defeating}
Ninon Eyrolles, Louis Goubin, and Marion Videau.
\newblock {Defeating {MBA}-based Obfuscation}.
\newblock In {\em ACM Workshop on Software PROtection (SPRO)}, 2016.

\bibitem{obf-vmbased}
Hui Fang, Yongdong Wu, Shuhong Wang, and Yin Huang.
\newblock {Multi-stage Binary Code Obfuscation using Improved Virtual Machine}.
\newblock In {\em International Conference on Information Security}, pages
  168--181. Springer, 2011.

\bibitem{feng2020neureduce}
Weijie Feng, Binbin Liu, Dongpeng Xu, Qilong Zheng, and Yun Xu.
\newblock {NeuReduce: Reducing Mixed Boolean-Arithmetic Expressions by
  Recurrent Neural Network}.
\newblock In {\em Conference on Empirical Methods in Natural Language
  Processing: Findings}, 2020.

\bibitem{gagnon2007software}
Michael~N Gagnon, Stephen Taylor, and Anup~K Ghosh.
\newblock {Software Protection through Anti-Debugging}.
\newblock {\em IEEE Security \& Privacy}, 5(3):82--84, 2007.

\bibitem{saturn}
Peter Garba and Matteo Favaro.
\newblock {{SATURN} -- Software Deobfuscation Framework Based On {LLVM}}.
\newblock In {\em ACM Workshop on Software PROtection (SPRO)}, 2019.

\bibitem{obf-controlflow}
Jun Ge, Soma Chaudhuri, and Akhilesh Tyagi.
\newblock {Control Flow based Obfuscation}.
\newblock In {\em ACM Workshop on Digital Rights Management}. ACM, 2005.

\bibitem{guinet2016arybo}
Adrien Guinet, Ninon Eyrolles, and Marion Videau.
\newblock {{Arybo}: Manipulation, Canonicalization and Identification of Mixed
  Boolean-Arithmetic Symbolic Expressions}.
\newblock In {\em {GreHack Conference}}, 2016.

\bibitem{gulwani2010dimensions}
Sumit Gulwani.
\newblock {Dimensions in Program Synthesis}.
\newblock In {\em International ACM SIGPLAN Symposium on Principles and
  Practice of Declarative Programming}, 2010.

\bibitem{radamsa}
Aki Helin.
\newblock {Radamsa: A General-purpose Fuzzer}.
\newblock \url{https://github.com/aoh/radamsa}.

\bibitem{intelpin}
{Intel Corporation}.
\newblock {Pin -- A Dynamic Binary Instrumentation Tool}.
\newblock
  \url{https://software.intel.com/en-us/articles/pin-a-dynamic-binary-instrumentation-tool}.

\bibitem{ollvm}
Pascal Junod, Julien Rinaldini, Johan Wehrli, and Julie Michielin.
\newblock {Obfuscator-{LLVM} -- Software Protection for the Masses}.
\newblock In {\em ACM Workshop on Software PROtection (SPRO)}, 2015.

\bibitem{kalysch2017vmattack}
Anatoli Kalysch, Johannes G{\"o}tzfried, and Tilo M{\"u}ller.
\newblock {VMAttack: Deobfuscating Virtualization-based Packed Binaries}.
\newblock In {\em Availability, Reliability and Security (ARES)}, 2017.

\bibitem{kinder2012towards}
Johannes Kinder.
\newblock {Towards Static Analysis of Virtualization-Obfuscated Binaries}.
\newblock In {\em IEEE Working Conference on Reverse Engineering (WCRE)}, 2012.

\bibitem{klees2018evalfuzz}
George Klees, Andrew Ruef, Benji Cooper, Shiyi Wei, and Michael Hicks.
\newblock {Evaluating Fuzz Testing}.
\newblock In {\em ACM Conference on Computer and Communications Security
  (CCS)}, 2018.

\bibitem{klint1981interpretation}
Paul Klint.
\newblock {Interpretation Techniques}.
\newblock {\em Software, Practice and Experience}, 11(9):963--973, 1981.

\bibitem{kroening2008decision}
Daniel Kroening and Ofer Strichman.
\newblock {\em {Decision Procedures}}.
\newblock Springer, 2016.

\bibitem{lee2018vodka}
Jae-Yung Lee, Jae~Hyuk Suk, and Dong~Hoon Lee.
\newblock {VODKA: Virtualization Obfuscation Using Dynamic Key Approach}.
\newblock In {\em International Workshop on Information Security Applications},
  2018.

\bibitem{liang2017deobfuscation}
Mingyue Liang, Zhoujun Li, Qiang Zeng, and Zhejun Fang.
\newblock {Deobfuscation of Virtualization-Obfuscated Code Through Symbolic
  Execution and Compilation Optimization}.
\newblock In {\em International Conference on Information and Communications
  Security}, 2017.

\bibitem{liu2021mbablast}
Binbin Liu, Junfu Shen, Jiang Ming, Qilong Zheng, Jing Li, and Dongpeng Xu.
\newblock {MBA-Blast: Unveiling and Simplifying Mixed Boolean-Arithmetic
  Obfuscation}.
\newblock In {\em USENIX Security Symposium}, 2021.

\bibitem{lu2020formal}
Weiyun Lu, Bahman Sistany, Amy Felty, and Philip Scott.
\newblock {Towards Formal Verification of Program Obfuscation}.
\newblock In {\em IEEE European Symposium on Security and Privacy Workshops
  (EuroS\&PW)}, 2020.

\bibitem{obf-selfmodify}
Matias Madou, Bertrand Anckaert, Patrick Moseley, Saumya Debray, Bjorn
  De~Sutter, and Koen De~Bosschere.
\newblock {Software Protection through Dynamic Code Mutation}.
\newblock In {\em International Workshop on Information Security Applications}.
  Springer, 2005.

\bibitem{menguy2021xyntia}
Gr{\'e}goire Menguy, S{\'e}bastien Bardin, Richard Bonichon, and Cauim de~Souza
  Lima.
\newblock {Search-Based Local Black-Box Deobfuscation: Understand, Improve and
  Mitigate}.
\newblock In {\em ACM Conference on Computer and Communications Security
  (CCS)}, 2021.

\bibitem{z3-homepage}
{Microsoft Research}.
\newblock {The Z3 Theorem Prover}.
\newblock \url{https://github.com/Z3Prover/z3}.

\bibitem{nakanishi2020intertwining}
Fukutomo Nakanishi, Giulio~De Pasquale, Daniele Ferla, and Lorenzo Cavallaro.
\newblock {Intertwining {ROP} Gadgets and Opaque Predicates for Robust
  Obfuscation}.
\newblock {\em CoRR}, abs/2012.09163, 2020.

\bibitem{ollivier2019howto}
Mathilde Ollivier, S\'{e}bastien Bardin, Richard Bonichon, and Jean-Yves
  Marion.
\newblock {How to Kill Symbolic Deobfuscation for Free (or: Unleashing the
  Potential of Path-Oriented Protections)}.
\newblock In {\em Annual Computer Security Applications Conference (ACSAC)},
  2019.

\bibitem{themida}
{Oreans Technologies}.
\newblock {Themida -- Advanced Windows Software Protection System}.
\newblock \url{https://www.oreans.com/Themida.php}.

\bibitem{obf-probfuscation}
Andre Pawlowski, Moritz Contag, and Thorsten Holz.
\newblock {Probfuscation: An Obfuscation Approach using Probabilistic Control
  Flows}.
\newblock In {\em Detection of Intrusions and Malware, and Vulnerability
  Assessment (DIMVA)}, 2016.

\bibitem{superoperators}
Todd~A. Proebsting.
\newblock {Optimizing an {ANSI} {C} Interpreter with Superoperators}.
\newblock In {\em ACM Symposium on Principles of Programming Languages (POPL)},
  1995.

\bibitem{rolles2009unpacking}
Rolf Rolles.
\newblock {Unpacking Virtualization Obfuscators}.
\newblock In {\em USENIX Workshop on Offensive Technologies (WOOT)}, 2009.

\bibitem{salwan2018symbolic}
Jonathan Salwan, S{\'e}bastien Bardin, and Marie-Laure Potet.
\newblock {Symbolic Deobfuscation: From Virtualized Code Back to the Original}.
\newblock In {\em Detection of Intrusions and Malware, and Vulnerability
  Assessment (DIMVA)}, 2018.

\bibitem{sarwar2013effectiveness}
Golam Sarwar, Olivier Mehani, Roksana Boreli, and Dali Kaafar.
\newblock {On the Effectiveness of Dynamic Taint Analysis for Protecting
  against Private Information Leaks on {Android}-based Devices}.
\newblock In {\em International Conference on Security and Cryptography
  (SECRYPT)}, 2013.

\bibitem{salwan2015triton}
Florent Saudel and Jonathan Salwan.
\newblock {Triton: A Dynamic Symbolic Execution Framework}.
\newblock In {\em Symposium sur la s{\'{e}}curit{\'{e}} des technologies de
  l'information et des communications (SSTIC)}, 2015.

\bibitem{schwartz2010all}
Edward~J Schwartz, Thanassis Avgerinos, and David Brumley.
\newblock {All You Ever Wanted to Know About Dynamic Taint Analysis and Forward
  Symbolic Execution (But Might Have Been Afraid to Ask)}.
\newblock In {\em IEEE Symposium on Security and Privacy}, 2010.

\bibitem{sharif2009automatic}
Monirul Sharif, Andrea Lanzi, Jonathon Giffin, and Wenke Lee.
\newblock {Automatic Reverse Engineering of Malware Emulators}.
\newblock In {\em IEEE Symposium on Security and Privacy}, 2009.

\bibitem{sharif2008impeding}
Monirul~I Sharif, Andrea Lanzi, Jonathon~T Giffin, and Wenke Lee.
\newblock {Impeding Malware Analysis Using Conditional Code Obfuscation}.
\newblock In {\em Symposium on Network and Distributed System Security (NDSS)},
  2008.

\bibitem{securom}
{Sony DADC}.
\newblock {{SecuROM} Software Protection}.
\newblock \url{https://www2.securom.com/Digital-Rights-Management.68.0.html}.

\bibitem{suk2020vcf}
Jae~Hyuk Suk and Dong~Hoon Lee.
\newblock {VCF: Virtual Code Folding to Enhance Virtualization Obfuscation}.
\newblock {\em IEEE Access}, 8, 2020.

\bibitem{llvm_homepage}
{The LLVM Project}.
\newblock {The {LLVM} Compiler Infrastructure}.
\newblock \url{https://llvm.org/}.

\bibitem{vanegue2012smt}
Julien Vanegue, Sean Heelan, and Rolf Rolles.
\newblock {{SMT} Solvers in Software Security}.
\newblock In {\em USENIX Workshop on Offensive Technologies (WOOT)}, 2012.

\bibitem{rc4implementation}
Robin Verton.
\newblock {RC4}.
\newblock \url{https://gist.github.com/rverton/a44fc8ca67ab9ec32089}, 2015.

\bibitem{libdvdcss}
{VideoLAN}.
\newblock {libdvdcss}.
\newblock \url{https://www.videolan.org/developers/libdvdcss.html}.

\bibitem{vlc}
{VideoLAN}.
\newblock {VLC} media player.
\newblock \url{https://www.videolan.org/}.

\bibitem{vmprotect}
{VMProtect Software}.
\newblock {VMProtect Software}.
\newblock \url{https://vmpsoft.com/}.

\bibitem{wang2011linear}
Zhi Wang, Jiang Ming, Chunfu Jia, and Debin Gao.
\newblock {Linear Obfuscation to Combat Symbolic Execution}.
\newblock In {\em European Symposium on Research in Computer Security
  (ESORICS)}, 2011.

\bibitem{weiser1981program}
Mark Weiser.
\newblock {Program Slicing}.
\newblock In {\em International Conference on Software Engineering (ICSE)},
  1981.

\bibitem{willsey2021egg}
Max Willsey, Chandrakana Nandi, Yisu~Remy Wang, Oliver Flatt, Zachary Tatlock,
  and Pavel Panchekha.
\newblock {egg: Fast and Extensible Equality Saturation}.
\newblock In {\em ACM Symposium on Principles of Programming Languages (POPL)},
  2021.

\bibitem{obf-general}
Gregory Wroblewski.
\newblock {General Method of Program Code Obfuscation}.
\newblock In {\em International Conference on Software Engineering Research and
  Practice (SERP)}, 2002.

\bibitem{xu2021boosting}
Dongpeng Xu, Binbin Liu, Weijie Feng, Jiang Ming, Qilong Zheng, Jing Li, and
  Qiaoyan Yu.
\newblock {Boosting SMT Solver Performance on Mixed-bitwise-arithmetic
  Expressions}.
\newblock In {\em ACM SIGPLAN Conference on Programming Language Design and
  Implementation (PLDI)}, 2021.

\bibitem{xu2018vmhunt}
Dongpeng Xu, Jiang Ming, Yu~Fu, and Dinghao Wu.
\newblock {VMHunt: A Verifiable Approach to Partially-virtualized Binary Code
  Simplification}.
\newblock In {\em ACM Conference on Computer and Communications Security
  (CCS)}, 2018.

\bibitem{xu2018manufacturing}
Hui Xu, Yangfan Zhou, Yu~Kang, Fengzhi Tu, and Michael Lyu.
\newblock {Manufacturing Resilient Bi-Opaque Predicates against Symbolic
  Execution}.
\newblock In {\em Conference on Dependable Systems and Networks (DSN)}, 2018.

\bibitem{xue2018exploiting}
Chao Xue, Zhanyong Tang, Guixin Ye, Guanghui Li, Xiaoqing Gong, Wei Wang,
  Dingyi Fang, and Zheng Wang.
\newblock {Exploiting Code Diversity to Enhance Code Virtualization
  Protection}.
\newblock In {\em International Conference on Parallel and Distributed
  Systems}, 2018.

\bibitem{yadegari2014bit}
Babak Yadegari and Saumya Debray.
\newblock {Bit-level Taint Analysis}.
\newblock In {\em IEEE Conference on Source Code Analysis and Manipulation},
  2014.

\bibitem{yadegari2015symbolic}
Babak Yadegari and Saumya Debray.
\newblock {Symbolic Execution of Obfuscated Code}.
\newblock In {\em ACM Conference on Computer and Communications Security
  (CCS)}, 2015.

\bibitem{yadegari2015generic}
Babak Yadegari, Brian Johannesmeyer, Ben Whitely, and Saumya Debray.
\newblock {A Generic Approach to Automatic Deobfuscation of Executable Code}.
\newblock In {\em IEEE Symposium on Security and Privacy}, 2015.

\bibitem{yang2011finding}
Xuejun Yang, Yang Chen, Eric Eide, and John Regehr.
\newblock {Finding and Understanding Bugs in {C} Compilers}.
\newblock In {\em ACM SIGPLAN Conference on Programming Language Design and
  Implementation (PLDI)}, 2011.

\bibitem{afl}
Micha\l{} Zalewski.
\newblock {American Fuzzy Lop}.
\newblock \url{http://lcamtuf.coredump.cx/afl/}.

\bibitem{zhou2007information}
Yongxin Zhou, Alec Main, Yuan~X Gu, and Harold Johnson.
\newblock {Information Hiding in Software with Mixed Boolean-Arithmetic
  Transforms}.
\newblock In {\em International Workshop on Information Security Applications},
  2007.

\end{thebibliography}

\subsection{Tigress}\label{sec:appendix:tigress}

\lstset{ 
  backgroundcolor=\color{white},
  basicstyle=\ttfamily\footnotesize,
  breakatwhitespace=false,
  breaklines=true,
  captionpos=b,
  title=\lstname
}

\begin{figure}[htb]
\begin{lstlisting}[
caption={Configuration of \tigress used to generate obfuscated samples. To guarantee randomness and diversity, we provided a unique seed per instance.},
label={lst:tigress_configuration}
]
    --Environment=x86_64:Linux:Clang:9.0 \
    --Seed=<UNIQUE_SEED> \
    --Transform=InitOpaque \
        --InitOpaqueStructs=list \
        --Functions=target_function \
    --Transform=InitImplicitFlow \
        --InitImplicitFlowHandlerCount=1 \
        --InitImplicitFlowKinds=bitcopy_signal \
        --Functions=target_function \
    --Transform=Virtualize \
        --Functions=target_function \
        --VirtualizeDispatch=indirect \
        --VirtualizeOptimizeBody=true \
        --VirtualizeOptimizeTreeCode=true \
        --VirtualizeOperands=registers \
        --VirtualizeSuperOpsRatio=2.0 \
        --VirtualizeMaxMergeLength=12 \
        --VirtualizeImplicitFlowPC=PCUpdate \
        --VirtualizeImplicitFlow="(single bitcopy_signal)" \
    --Transform=EncodeArithmetic \
        --Functions=target_function \
        --EncodeArithmeticKinds=integer
\end{lstlisting}
\end{figure}

To achieve comparability, we configure \tigress to resemble our approach in terms of VM architecture, superoperators, and arithmetic encodings. The detailed configuration of \tigress is depicted in Listing~\ref{lst:tigress_configuration}. In short, we configure a VM architecture with an inlined dispatcher (featuring direct threaded code), an upper bound for superoptimization with a depth of $12$ (in line with \loki's upper recursion bound of $12$), and allow optimizations to decrease overhead. Additionally, we use MBAs (called \emph{encode arithmetic}) to harden the code further and employ Tigress' anti-taint feature. When creating binaries with \tigress, we assign each instance a unique seed to produce a diverse set of obfuscated binaries.

\subsection{MBA Overhead}\label{sec:appendix:mba_overhead}
We analyze the overhead of our MBAs further.

\begin{experiment}[MBA Overhead]\label{experiment:benchmarking:mba_bounds}
To further quantify the overhead of MBAs in terms of additional instructions, we recursively apply MBAs for a bound $r \in\{0, 10, 20, 30, 40, 50\}$. For each $r$, we generate $100$ obfuscated binaries and average the number of instructions over \loki's 512 handlers.
\end{experiment}

\begin{table}[htb]
\centering
\caption{Average number of assembly instructions per handler for various recursion bounds.}
\resizebox{\linewidth}{!}{
\begin{tabularx}{1.17\linewidth}{lRRRRRR}
\toprule
\textbf{Bound} & 0 & 10 & 20 & 30 & 40 & 50 \\
\textbf{Instructions} & 111 & 164 & 217 & 269 & 321 & 372 \\
\bottomrule
\end{tabularx}
}%
\label{tab:mba_overhead} %
\end{table}

As can be seen in Table~\ref{tab:mba_overhead}, increasing the recursion bound by $10$ adds \textasciitilde$52$ new assembly instructions.
This experiment visualizes the trade-off between achieving a sufficient level of protection and overhead. To strike a good balance, we randomly choose a recursion bound between $20$ and $30$ in our implementation, resulting in \textasciitilde$222$ instructions per handler (cf. Table~\ref{tab:intro:commercial_vm_stats}). Given the MBAs' importance with regard to deterring an attacker, we consider their overhead acceptable.

\subsection{Symbolic Execution for Different MBA Bounds}\label{sec:appendix:se_mba_bounds}

In the following, we analyze the impact of \loki's recursive rewriting bounds (chosen randomly between 20 and 30 by our prototype) on its security.

\begin{experiment}[Impact of MBA bounds on Security]
We create $1,000$ binaries with handlers as described in Section~\ref{sec:evaluaton:syntactic_attacks} for each recursive MBA rewriting bound $r \in\{0, 5, 10, 15, 20, 25, 30, 35, 40, 45, 50, 55\}$. 
Similar to \ex{experiment:symbolic_execution}, we do this for handlers of semantic depth $3$ and $5$.
We then evaluate for each handler, whether a dynamic attacker using \miasm's symbolic execution engine can simplify the MBAs to their original core semantics (given a timeout of one hour).
\end{experiment}

\begin{figure}[htb]
\centering
    \includegraphics[width=.9\linewidth]{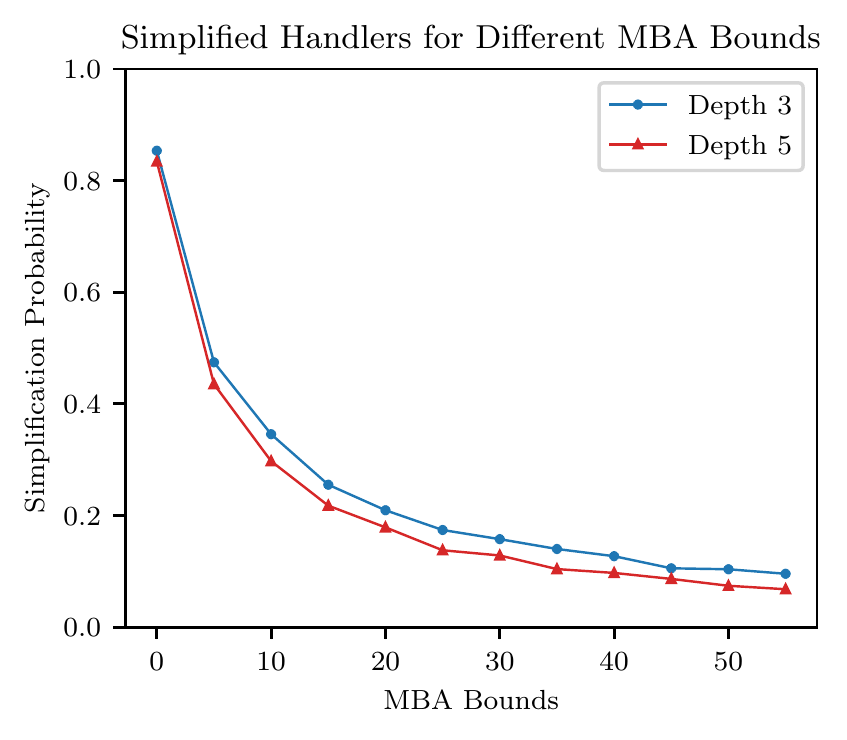}
    \caption{The probability of simplifying a handler per MBA bound. An MBA bound refers to the number of times MBAs were recursively applied, 0 indicates no MBA were applied.}%
    \label{fig:se_mba_bounds_appendix}%
\end{figure}

The results are plotted in Figure~\ref{fig:se_mba_bounds_appendix}. 
For all bounds, it is visible that a higher semantic depth correlates with less handlers simplified (on average, the distance of simplified handlers between depth 3 and depth 5 is $3.3\%$). This confirms that superoperators have beneficial synergy effects as they cause core semantics to have a higher semantic depth.
If users of \loki desire a higher security than our prototype provides, they can set a bound of $55$ to reduce the number of handlers simplified to $9.56\%$ (depth 3) or $6.79\%$ (depth $5$), respectively. However, the higher level of security comes at the cost of increased overhead, both in terms of space and runtime (\cf Experiments~\ref{experiment:overhead}~and~\ref{experiment:benchmarking:mba_bounds}).

\end{document}